\documentclass[12pt]{iopart}
\usepackage{amssymb}
\usepackage{graphicx}% Include figure files
\usepackage{dcolumn}% Align table columns on decimal point
\usepackage{bm}% bold math
\usepackage{epsfig}
\newcommand{\beq}{\begin{equation}}
\newcommand{\eeq}{\end{equation}}
\newcommand{\beqa}{\begin{eqnarray}}
\newcommand{\eeqa}{\end{eqnarray}}
\newcommand{\ba}{\begin{array}}
\newcommand{\ea}{\end{array}}
\usepackage{color}

\begin{document}

\title{Rabi-Josephson oscillations and self-trapped dynamics
in atomic junctions with two bosonic species}
\author{G. Mazzarella$^{1}$, B. Malomed$^{2}$, L. Salasnich$^{1,3}$,
M. Salerno$^{4}$, and F. Toigo$^{1}$}
\address{$^{1}$Dipartimento di Fisica \textquotedblleft Galileo
Galilei\textquotedblright\ and CNISM, Universit\`{a} di Padova, Via Marzolo
8, 35122 Padova, Italy$^{1}$ \\
$^{2}$Department of Physical Electronics, Faculty of Engineering, Tel Aviv
University, Tel Aviv 69978, Israel \\
$^{3}$INO-CNR, via G. Sansone 1, 50019 Sesto Fiorentino, Italy\\
$^{4}$Dipartimento di Fisica ``E.R. Caianiello'', CNISM and INFN - Gruppo
Collegato di Salerno, Universit\`a di Salerno, Via Ponte don Melillo, 84084
Fisciano (SA), Italy}
\date{\today}

\begin{abstract}
We investigate the dynamics of two-component Bose-Einstein
condensates (BECs), composed of atoms in two distinct hyperfine
states, which are linearly coupled by two-photon Raman transitions.
The condensate is loaded into a double-well potential (DWP). A
variety of dynamical behaviors, ranging from regular Josephson
oscillations, to mixed Rabi-Josephson oscillations and to regimes
featuring an increasing complexity, are described in terms of a
reduced Hamiltonian system with four degrees of freedoms, which are
the numbers of atoms in each component in the left and right
potential wells, whose canonically conjugate variables are phases of
the corresponding wave functions. Using the system with the four
degrees of freedom, we study the dynamics of fractional imbalances
of the two bosonic components, and compare the results to direct
simulations of the Gross-Pitaevskii equations (GPEs) describing the
bosonic mixture. We perform this analysis when the fractional imbalance
oscillates around a zero-time averaged value and in the self-trapping regime
as well.
\end{abstract}

\maketitle

%%%%%%%%%%%%%%%%%%

\section{Introduction}

The study of Josephson oscillations and self-trapping both with a
single bosonic component \cite{smerzi} and in bosonic binary mixtures \cite
{myatt,lobo,xu,satja,diaz1,ajj,diaz2,Amherst} trapped in double-well potentials (DWPs) has
attracted much interest in the context of the current work on ultracold
quantum gases. A specific ramification of this topic corresponds
to the situation
in which the two components of the mixture are different hyperfine states
of the
same bosonic atom \cite{myatt,lobo,diaz1}, which may be linearly coupled by
an external resonant field \cite{burnett}. This setting suggests a
possibility to study the interplay between Josephson and Rabi
oscillations, the latter being induced by the linear interconversion
between the
components \cite{cornell,Rabi-oscill}. This subject was considered in
several earlier works \cite{yasunaga,martinez,hai}. In particular,
in Ref. \cite{yasunaga} the authors analyzed a crossover between
the Josephson and Rabi
dynamics, using a nonstationary model (with a linearly growing
magnetic field), which, in the Josephson limit, was reduced to a
system of two degrees of freedom. In Ref. \cite{martinez}, a
two-degrees-of-freedom model was used too, with the objective of
studying the quasiparticles' spectrum in the symmetry-broken ground
state. As suggested in Ref. \cite{pezze}, there exists a possibility
to distinguish between the Rabi and Josephson regimes by
considering a beam-splitter model based on a nonstationary DWP.
%The
%onset of dynamical chaos in the two-components system was studied
%too, within the framework of a model with $1.5$ degrees of freedom,
%see Ref. \cite{hai} and references therein.
%{\color{red} Incidentally, note that an example of system
%with $1.5$ degrees of freedoms is provided by a planar rigid pendulum driven by an ac force.
%One degree of freedom is obviously accounted
%by the pendulum angle, say $\phi$, and associated angular velocity, $\dot{\phi}$, while
%the half degree of freedom comes from the time, $t$, regarded as coordinate, say
%$y=t$,  with a constant associated momentum; this gives for the
%$y$ variable only $1/2$ degree instead than $1$. In general, in this way one can
%always reduce a $f$ degrees of freedom non autonomus system  to a $f+1/2$ degrees
%of freedom autonomous one, a fact  which may be useful especially when the
%time dependence is periodic.}
An experimental implementation of
internal bosonic Josephson junctions with Rubidium
spinor Bose-Einstein condensate has been recently considered
by Zibold {\it et al.} \cite{obert} in connection with
bifurcations occurring at the transition from Rabi to Josephson dynamics.

The natural minimum basis for the analysis of the Rabi-Josephson
oscillations in two-component systems in the DWP should include
\emph{four} degrees of freedom \cite{Amherst}.
%while the previously
%studied reductions
%to two and $1.5$ degrees of freedom cannot comprise generic situations.
The objective of this work is to analyze physically relevant dynamical regimes
within the framework of the minimal system. The predictions will be verified via
the comparison to direct simulations of the underlying Gross-Pitaevskii
equations (GPEs).

The paper is organized as follows. The model is described in Section II, and
the finite-mode approximation is derived in Section III, where we also verify
its accuracy by comparison to direct simulations of the underlying
GPEs. In Section IV, we report the main results of
the work, which demonstrate the interplay between the Josephson and Rabi
oscillations, concluding that the oscillations become characterized by an high
degree of complexity with the increase of the strength of Rabi coupling.
In order
to enlarge and complete the analysis presented in Ref. \cite{diaz2},
the study of the self-trapping regime is faced as well. In fact, in Section V - within the
self-trapping regime - we discuss the accuracy of our model following the
same path as in Sec. IV; we comment about the influence of the linear coupling
constant on the onset of the self-trapping dynamics. The paper is concluded by
Section VI.

\section{The model}

We consider a binary Bose-Einstein condensate of two
different species (with index $1$ and $2$) of repulsively interacting
bosons. The condensate is trapped in a DWP, which can be produced, for
example, by a far off-resonance laser barrier separating each component into
two regions, $L$ (left) and $R$ (right). These components may be , for
example, two distinct hyperfine states, $|F=2,m_{F}=1\rangle $ and
$|F=1,m_{F}=-1\rangle $, of $^{87}${Rb} \cite{myatt,lobo}. Note that since
$|\Delta m_{F}|=2$ one needs two photons to couple the two levels.
A weak external
magnetic field gives rise to a small difference, $\hbar \omega _{0}$,
between the energy levels of these states. Two-photon Raman transitions
between the levels, characterized by Rabi frequency $\Omega $, can be
induced by a laser beam of frequency $\omega _{d}$, with detuning $\delta
\equiv \omega _{d}-\omega _{0}$. Using the rotating-wave approximation
(e.g., neglecting high-frequency terms in the atom-field interaction), in
the mean-field approximation macroscopic wave functions
$\Psi _{n}(\mathbf{r},t)$ ($n=1,2$) of the two components of the condensates
obey the system of
coupled Gross-Pitaevskii equations (GPEs) \cite{cornell,two-photon}:
\begin{eqnarray}
i\hbar \frac{\partial \Psi _{n}}{\partial t} &=&-\frac{\hbar ^{2}}{2m}\nabla
^{2}\Psi _{n}+[V_{\mathrm{trap}}^{(n)}(\mathbf{r})+\frac{(-1)^{n}}{2}\,\hbar
\,\delta +g_{n}|\Psi _{n}|^{2}  \nonumber  \label{GPE} \\
&+&g_{12}|\Psi _{3-n}|^{2}]\Psi _{n}+\frac{\Omega }{2}\Psi _{3-n},
\end{eqnarray}%
where $V_{\mathrm{trap}}(\mathbf{r})$ is the trapping potential and $\Psi
_{n}(\mathbf{r},t)$ is subject to the normalization condition,
\begin{equation}
\int d^{3}\mathbf{r}\,|\Psi _{n}(\mathbf{r},t)|^{2}=N_{n}(t),
\label{normalizationwf}
\end{equation}%
with $N_{n}(t)$ the number of bosons of the $n$-th species.
Similarly, $m$,
$a_{n}$, and $g_{n}=4\pi \hbar ^{2}a_{n}/m_{n}$ denote, respectively, the
atomic mass, $s$-wave scattering length, and intra-species nonlinearity
coefficient of the $n$-th species (the atomic mass is common for both
species). The constant accounting for the linear interconversion between the
bosonic components is expressed in terms of the respective Rabi frequency,
$\Omega $. Finally, $g_{12}=2\pi \hbar ^{2}a_{12}/m$ is the coefficient
accounting for the nonlinear interaction between the species, $a_{12}$ being
the respective $s$-wave scattering length. Notice that the total
number of particles $N_1(t)+N_2(t)$ is a conserved quantity.
In the following, we consider
both $g_{n}$ and $g_{12}$ as free parameters, due to the possibility to
change the scattering lengths by means of the Feshbach-resonance technique,
see, e.g., Ref. \cite{Ueda} and references therein.

Equations (\ref{GPE}) can also be derived in a different physical setting,
by assuming that the two hyperfine states may be coupled by an external ac
magnetic field $B\cos (\omega t)$ of frequency $\omega =\omega _{0}-\delta$.
In this case, the linear coupling term in Eq. (\ref{GPE}) corresponds to
the Rabi frequency $\Omega =\mathbf{\mu }\cdot \mathbf{B}/\hbar $, where
$\mu $ is the dipole matrix element for the transition between the two
hyperfine states \cite{burnett}.

The trapping potential for both components is taken to be the superposition
of a strong harmonic confinement in the transverse ($x$,$y$) plane and of a
DWP in the axial ($z$) direction, i.e.,
\begin{equation}
V_{\mathrm{trap}}^{(n)}(\mathbf{r})=\frac{1}{2}m\omega_{n}^{2}
(x^{2}+y^{2})+V_{\mathrm{DWP}}(z) \; .
\label{trap}
\end{equation}

We proceed by writing the Lagrangian associated to the GPEs (\ref{GPE}),
\begin{eqnarray}
L &=&\int d^{3}\mathbf{r}\,\bigg(\bigg[\sum_{n=1,2}\bar{\Psi}_{n}\big(i\hbar
\frac{\partial }{\partial t}+\frac{\hbar ^{2}}{2m_{n}}\nabla ^{2}\big)\Psi
_{n}  \nonumber  \label{lagrangian} \\
&-&\big(V_{\mathrm{trap}}^{(n)}(\mathbf{r})|\Psi _{n}|^{2}+
\frac{(-1)^{n}\,\hbar \,\delta }{2}|\Psi _{n}|^{2}+\frac{g_{n}}{2}
|\Psi _{n}|^{4}\big)\bigg]  \nonumber \\
&&-(\Omega /2)\left( \bar{\Psi}_{1}\Psi _{2}+\Psi _{1}\bar{\Psi}_{2}\right)
-g_{12}|\Psi _{1}|^{2}|\Psi _{2}|^{2}\bigg)\;,
\end{eqnarray}%
where $\bar{\Psi}_{n}$ stands for the complex conjugate of $\Psi _{n}$. To
derive, at first, the 1D approximation, we adopt the usual ansatz
\begin{equation}
\Psi _{n}(x,y,z,t)=\frac{1}{\sqrt{\pi }a_{\bot ,n}}\exp
\left( -\frac{x^{2}+y^{2}}{2a_{\bot ,n}^{2}}\right) f_{n}(z,t)\;,
\label{ansatz}
\end{equation}%
where
$a_{\bot,n}=\sqrt{\hbar/\left(m\omega_{n}\right)}$
are the respective transverse-confinement radii, with the 1D wave functions
$f_{n}(z,t)$ obeying normalization conditions
$\int dz|f_{n}(z,t)|^{2}=N_{n}(t)$.
Note that the factorized ansatz (\ref{ansatz}) is
valid under the strong transverse confinement, \textit{viz}., when
$g_{n}|f_{n}|^{2}/4\pi a_{\bot ,n}^{2}\ll \hbar \omega _{n}$
\cite{salasnichvecsol}.
By inserting the ansatz (\ref{ansatz}) into the Lagrangian
(\ref{lagrangian}) and
performing the integration in the transverse plane, we derive the effective
Lagrangian for the 1D wave functions:
\begin{eqnarray}
L_{\mathrm{1D}} &=&\int dz\,\bigg(\bigg[\sum_{n=1,2}
\bar{f}_{n}\big(i\hbar \frac{\partial }{\partial t}+\frac{\hbar ^{2}}{2m}
\frac{\partial ^{2}}{\partial z^{2}}\big)f_{n}  \nonumber
\label{effectivelagrangian} \\
&-&\left[ \epsilon _{n}+V_{\mathrm{DWP}}(z)+\frac{(-1)^{n}\,\hbar \,\delta }
{2}\right] |f_{n}|^{2}-\frac{\tilde{g}_{n}}{2}|f_{n}|^{4}\bigg]  \nonumber \\
&-&\left( \tilde{\Omega} /2\right)
\left( \bar{f}_{1}f_{2}+f_{1}\bar{f}_{2}\right) -
\tilde{g}_{12}|f_{1}|^{2}|f_{2}|^{2}\bigg)\;,
\nonumber
\end{eqnarray}
where the following constants are introduced:
\beq
\epsilon_{n} = (1/2)
\left[ \hbar ^{2}/\left(m a_{n,\bot}^2\right)
+m\omega _{n}^{2}a_{n,\bot}^2\right] \; ,
\eeq
\beq
\tilde{g}_{n} = {g_{n}/}\left( 2\pi a_{\bot ,n}^{2}\right) \; ,
\eeq
\beq
\tilde{\Omega} = 2a_{\bot,1}a_{\bot,2}
(a_{\bot,1}^{2}+a_{\bot,2}^{2})^{-1} \Omega \; ,
\eeq
and
\beq
\tilde{g}_{12} = \left( g_{12}/\pi \right)
\left( a_{\bot ,1}^{2}+a_{\bot ,2}^{2}\right)^{-1} \; .
\eeq
By varying $L_{\mathrm{1D}}$ with respect to
$\bar{f}_{n}$, we derive the effective 1D GPEs,
\begin{eqnarray}
i\hbar \frac{\partial f_{n}}{\partial t} &=&-\frac{\hbar ^{2}}{2m_{n}}
\frac{\partial ^{2}f_{n}}{\partial z^{2}}
+[\epsilon _{n}+V_{\mathrm{DWP}}(z)+
\frac{(-1)^{n}\,\hbar \,\delta }{2}
\nonumber
\label{1dGPE}
\\
&+&\tilde{g}_{n}|f_{n}|^{2}+\tilde{g}_{12}|f_{3-m}|^{2}]f_{n}+
\left( \tilde{\Omega}/2\right) f_{3-m}\;.
\end{eqnarray}

\section{The finite-mode system}

To approximate the dynamics by a finite-mode approximation, we make use of
the two-mode decomposition for each wave function, as originally introduced
in Ref. \cite{milburn}:
\begin{equation}
f_{n}(z,t)=\psi _{n}^{L}(t)\phi _{n}^{L}(z)
+\psi _{n}^{R}(t)\phi_{n}^{R}(z)\;.  \label{f}
\end{equation}%
The orthonormal real functions $\phi_{n}^{\alpha }(z)$ are constructed
according to the same path as in Ref. \cite{ajj} and as commented below.
These functions are localized in the left and in the right wells,
respectively ($\alpha =L,R$) \cite{ajj}, and
\begin{equation}
\psi_{n}^{\alpha }(t)\equiv \sqrt{N_{n}^{\alpha }(t)} \
e^{i\theta _{n}^{\alpha}(t)}\;,
\label{guess}
\end{equation}%
with the total number of particles in the $n$-th species being
$N_{n}^{L}(t)+N_{n}^{R}(t)=\left\vert \psi _{n}^{L}(t)\right\vert ^{2}
+\left\vert\psi _{n}^{R}(t)\right\vert ^{2}\equiv N_{n}(t)$.
As described in the Appendix,
we derive explicit evolution equations for the
temporal evolution of the \textit{fractional imbalances},
$z_{n}=(N_{n}^{L}-N_{n}^{R})/N_{n}$ and intra-species relative phases,
$\theta _{n}=\theta _{n}^{R}-\theta_{n}^{L}$:
%\begin{widetext}
\begin{eqnarray}
\dot{z}_{n} &=&-\frac{2(K_{n}-K_{c,n}N_{n})}{\hbar }\,\sqrt{1-z_{n}^{2}}%
\,\sin \theta _{n}+\frac{V_{n}N_{n}}{2\hbar }(1-z_{n}^{2})\sin 2\theta _{n}
\nonumber  \label{odesztheta} \\
&+&\frac{2}{\hbar }(V_{12}\sqrt{1-z_{3-n}^{2}}\,\cos \theta
_{3-n}+K_{c,12})N_{3-n}\sqrt{1-z_{n}^{2}}\,\sin \theta _{n}  \nonumber \\
&&\mp \frac{\tilde \Omega }{2\hbar }\sqrt{N_{1}N_{2}}
\bigg(\sqrt{(1+z_{1})(1+z_{2})}%
\sin \gamma _{L}-\sqrt{(1-z_{1})(1-z_{2})}\sin \gamma _{R}\bigg),
\nonumber
\\
\dot{\theta}_{n} &=&\frac{U_{n}-V_{n}}{\hbar }N_{n}z_{n}+\frac{%
2(K_{n}-K_{c,n}N_{n})}{\hbar }\,\frac{z_{n}\cos \theta _{n}}{\sqrt{%
1-z_{n}^{2}}}\nonumber\\
&-&\frac{V_{n}N_{n}}{2\hbar }\,z_{n}\cos 2\theta _{n}+\frac{%
U_{12}-V_{12}}{\hbar }N_{3-n}z_{3-n}  \nonumber \\
&-&\frac{2}{\hbar }\left[ V_{12}\sqrt{1-z_{3-n}^{2}}\,\cos \theta
_{3-n}+K_{c,12}\right] N_{3-n}\frac{z_{n}\cos \theta _{n}}
{\sqrt{1-z_{n}^{2}}%
}  \nonumber \\
&+&\frac{\tilde{\Omega}}{2\hbar }\sqrt{\frac{N_{3-n}}{N_{n}}}
\bigg(\sqrt{\frac{%
1+z_{3-n}}{1+z_{n}}}\cos \gamma _{L}-\sqrt{\frac{1-z_{3-n}}{1-z_{n}}}\cos
\gamma _{R}\bigg)\;.
\end{eqnarray}%
%\end{widetext}
Here the signs plus and minus pertain to $n=1$ and $2$,
respectively. The total numbers of particles $N_{n}$ of each
component and the respective phases, $\gamma _{\alpha }=\theta
_{1}^{\alpha }-\theta _{2}^{\alpha }$ (recall $\alpha =L,R$) evolve
according to:
%\begin{widetext}
\begin{eqnarray}
&&\dot{N_{n}}=\pm (-\frac{\tilde{\Omega}}{2\hbar }\sqrt{N_{1}N_{2}})\big(%
\sqrt{(1+z_{1})(1+z_{2})}\,\sin \gamma _{L}+\sqrt{(1-z_{1})(1-z_{2})}\,\sin
\gamma _{R}\big),
\nonumber
\label{nitetaalpha}
\\
&&\dot{\gamma}_{L}=\frac{1}{2\hbar }\bigg(%
N_{1}(U_{12}-U)(1+z_{1})-N_{2}(U_{12}-U)(1+z_{2})+\Delta E\bigg)\nonumber\\
&+&\frac{1}{\hbar}\bigg(K_{1}\sqrt{\frac{1-z_{1}}{1+z_{1}}}\cos \theta
_{1}-K_{2}\sqrt{\frac{%
1-z_{2}}{1+z_{2}}}\cos \theta _{2}\bigg)
\nonumber
\\
&-&\frac{\tilde{\Omega}}{2\hbar \sqrt{N_{1}N_{2}}}\bigg(\frac{%
N_{2}(1+z_{2})-N_{1}(1+z_{1})}{\sqrt{(1+z_{1})(1+z_{2})}}\bigg)\cos \gamma
_{L}\nonumber\\
&+&\frac{K_{c,12}}{\hbar }\big[N_{1}\sqrt{1-z_{1}^{2}}\cos \theta
_{1}-N_{2}\sqrt{1-z_{2}^{2}}\cos \theta _{2}\big]-
\nonumber
\\
&&\frac{V}{2\hbar }\bigg[N_{1}(1-z_{1})(2+\cos 2\theta
_{1})-N_{2}(1-z_{2})(2+\cos 2\theta _{2})\bigg]-  \nonumber \\
&&\frac{1}{\hbar }\bigg[K_{c,1}N_{1}(2+z_{1})+K_{c,12}N_{2}+V_{12}N_{2}\sqrt{%
1-z_{2}^{2}}\cos \theta _{2}\bigg]\sqrt{\frac{1-z_{1}}{1+z_{1}}}\cos \theta
_{1}+
\nonumber
\\
&&\frac{1}{\hbar }\bigg[K_{c,2}N_{2}(2+z_{2})+K_{c,12}N_{1}\nonumber\\
&+&V_{12}N_{1}\sqrt{%
1-z_{1}^{2}}\cos \theta _{1}\bigg]\sqrt{\frac{1-z_{2}}{1+z_{2}}}\cos \theta
_{2}+\frac{V_{12}}{2}\big(N_{1}(1-z_{1})-N_{2}(1-z_{2})\big),
\nonumber
\\
&&\dot{\gamma}_{R}=\frac{1}{2\hbar }\bigg(%
N_{1}(U_{12}-U)(1-z_{1})-N_{2}(U_{12}-U)(1-z_{2})+\Delta E\bigg)\nonumber\\
&+&\frac{1}{%
\hbar }\bigg(K_{1}\sqrt{\frac{1+z_{1}}{1-z_{1}}}\cos \theta _{1}-K_{2}\sqrt{\frac{%
1+z_{2}}{1-z_{2}}}\cos \theta _{2}\bigg)-
\nonumber
\\
&&\frac{\tilde{\Omega}}{\hbar \sqrt{N_{1}N_{2}}}\bigg(\frac{%
N_{2}(1-z_{2})-N_{1}(1-z_{1})}{\sqrt{(1-z_{1})(1-z_{2})}}\bigg)\cos \gamma
_{R}\nonumber\\
&+&\frac{K_{c,12}}{\hbar }\big[N_{1}\sqrt{1-z_{1}^{2}}\cos \theta
_{1}-N_{2}\sqrt{1-z_{2}^{2}}\cos \theta _{2}\big]-
\nonumber
\\
&&\frac{V}{2\hbar }\bigg[N_{1}(1+z_{1})(2+\cos 2\theta
_{1})-N_{2}(1+z_{2})(2+\cos 2\theta _{2})\bigg]-  \nonumber \\
&&\frac{1}{\hbar }\bigg[K_{c,1}N_{1}(2-z_{1})+K_{c,12}N_{2}+V_{12}N_{2}\sqrt{%
1-z_{2}^{2}}\cos \theta _{2}\bigg]\sqrt{\frac{1+z_{1}}{1-z_{1}}}\cos \theta
_{1}+
\nonumber
\\
&&\frac{1}{\hbar }\bigg[K_{c,2}N_{2}(2-z_{2})+K_{c,12}N_{1}\nonumber\\
&+&V_{12}N_{1}\sqrt{%
1-z_{1}^{2}}\cos \theta _{1}\bigg]\sqrt{\frac{1+z_{2}}{1-z_{2}}}\cos \theta
_{2}+\frac{V_{12}}{2}\big(N_{1}(1+z_{1})-N_{2}(1+z_{2})\big),
\nonumber \\
&&
\end{eqnarray}
%\end{widetext}
where $\Delta E\equiv E_{2}-E_{1}=\hbar \,\delta $. The latter
relation follows from the normalization of functions $\phi _{\alpha
}$ and the first equation from Eq. (\ref{parameters}). Obviously, in
the absence of Rabi coupling, i.e. $\tilde{\Omega}=0$, the total
number of particles is conserved in each component. In that case,
Eqs. (\ref{odesztheta}) reduce to the equations of motion for
coupled pendula derived in Ref. \cite{ajj}. %{\color{red}
We observe
that due to the nonlinearity associated to the intra- and
inter-species interactions, the system is nonintegrable also when
$\tilde \Omega=0$.
%} {\color{red} 
On the other hand, in the absence
of the Josephson coupling, i.e., when $K_{n}=0$, the dynamics will
be characterized by independent Rabi oscillations in each well; when
the Josephson coupling is finite, i.e., $K_{n} \neq 0$, and much
smaller than $\tilde \Omega$, the aforementioned single-well
independent Rabi oscillations will be weakly coupled by the
Josephson tunneling. Note that both when $K_{n}=0$ and $K_{n} \neq
0$, the Rabi interconversions will be deformed by nonlinear effects
due to the intra- and the inter-species interactions.
%}

In our calculations the axial double-well potential is given by \beq
\label{dwp} V_{DWP}(z)=-U_{0}\bigg[{\rm sech}^2(\frac{z+z_0}{{\cal
A}}) +{\rm sech}^2(\frac{z-z_0}{{\cal A}})\bigg] \eeq with \beq
U_{0}=\hbar \omega_{\bot} \big[1+{\rm sech}^2(\frac{2z_0}{{\cal A}})
\big]^{-1} \; , \eeq that is the combination of two
P\"{o}schl-Teller (PT) potentials separated by a potential barrier
the height of which may be changed by changing
 ${\cal A}$, and centered around the points $-z_0$ and $z_0$. The
the wave functions of the eigenvalues problem in the presence of the
only potential
$V_{\alpha}(z)$ ($\alpha=L,R$) are exactly known. In particular, the wave
function of the ground state is \cite{landau}
\begin{eqnarray}
\label{wfpt} &&\phi_{n}^{\alpha,PT}(z)=B\big[1-Tanh^{2}(\frac{z
\mp
z_0}{{\cal A}})\big]^{C_{n}/2} \nonumber\\
&& C_{n}=-\frac{1}{2}+\sqrt{\frac{2mU_0
{\cal A}^2}{\hbar^{2}}+\frac{1}{4}} \; .\end{eqnarray} If $\alpha=L$
($R$), the function $\phi_{n}^{\alpha,PT}(z)$ is centered around
$-z_0$ ($+z_0$), and it is the ground state wave function of the PT potential
centered around the point $-z_0$ ($z_0$). In Eq. (\ref{wfpt}) $B$,
equal for both
sides, ensures the normalization of the wave function.
The above functions $\phi_{n}^{L}(z)$ and  $\phi_{n}^{R}(z)$.
 Then, proceeding from the functions (\ref{wfpt}), $\phi_{L}(z)$
and $\phi_{R}(z)$ can be determined following the same perturbative
approach as in \cite{ajj}, where it is shown that, under certain
hypothesis, the aforementioned functions can be written
in terms of $\phi_{n}^{L,PT}(z)$
and$\phi_{n}^{R,PT}(z)$ given by (\ref{wfpt}). We get \cite{ajj}
%\begin{widetext}
\begin{eqnarray}
\label{finaldecomposition} &&
\phi_{n}^{L}(z)=\frac{1}{2}\bigg[(\frac{1}{\sqrt{1+s}}+\frac{1}
{\sqrt{1-s}})\phi_{n}^{L,PT}(z)+(\frac{1}{\sqrt{1+s}}-\frac{1}
{\sqrt{1-s}})\phi_{n}^{R,PT}(z)\bigg] \nonumber\\
&&
\phi_{n}^{R}(z)=\frac{1}{2}\bigg[(\frac{1}{\sqrt{1+s}}-\frac{1}
{\sqrt{1-s}})\phi_{n}^{L,PT}(z)+(\frac{1}{\sqrt{1+s}}+\frac{1}
{\sqrt{1-s}})\phi_{n}^{R,PT}(z)\bigg] \; ,
\nonumber \\
\end{eqnarray}
%\end{widetext}
where $s=\int_{-\infty}^{+\infty}dz\, \phi_{n}^{L,PT}(z)
\phi_{n}^{R,PT}(z)$.

The relations between the macroscopic parameters involved at
right hand sides of
Eqs. (\ref{odesztheta})-(\ref{nitetaalpha}) and the microscopic
parameters of the problem are reported in the Eq. (\ref{parameters})
of the Appendix. We observe, moreover, that to make the system fully
symmetric we also assume that $K_{1}=K_{2}\equiv K$, $%
U_{1}=U_{2}\equiv U$, $K_{c,1}=K_{c,2}\equiv K_{c}$,
and $V_{1}=V_{2}\equiv V $.

Obviously, in the absence of Rabi
coupling, i.e., $\tilde{\Omega}=0$, the total number of particles is
conserved in each component. In that case, Eqs. (\ref{odesztheta}) reduce to
the equations of motion for coupled pendula derived in Ref. \cite{ajj}.
Note, moreover, that in our calculations we have assumed that the
both the components feel the same harmonic potential
so that $\omega_1=\omega_2 \equiv \omega_{\bot}$.

\begin{figure}[h!]
\centerline{\includegraphics[width=4cm,clip]{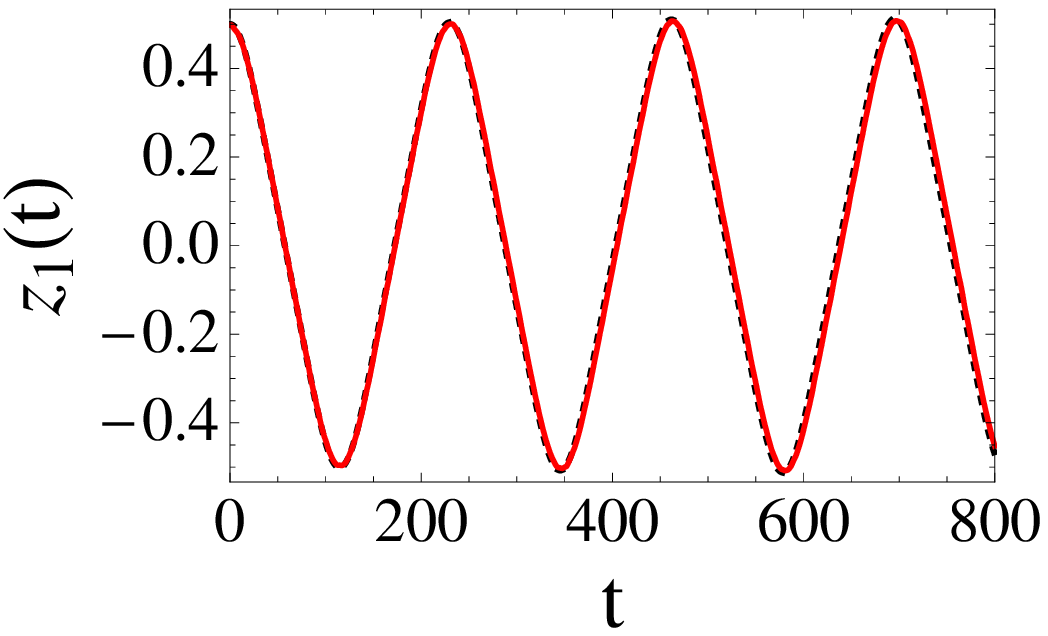}
\includegraphics[width=4cm,clip]{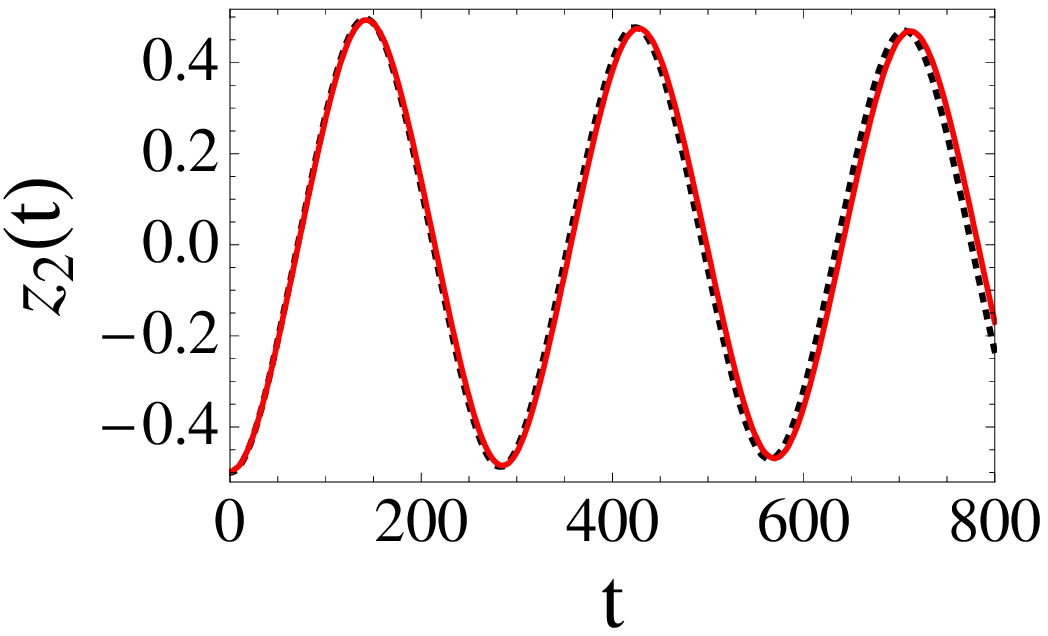}}
\centerline{\includegraphics[width=4cm,clip]{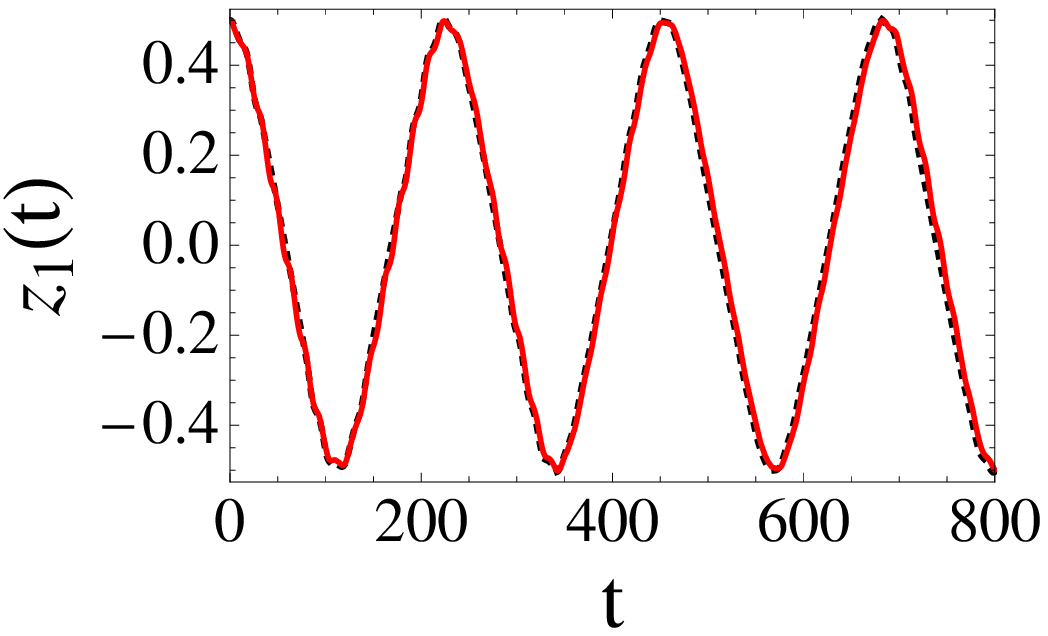}
\includegraphics[width=4cm,clip]{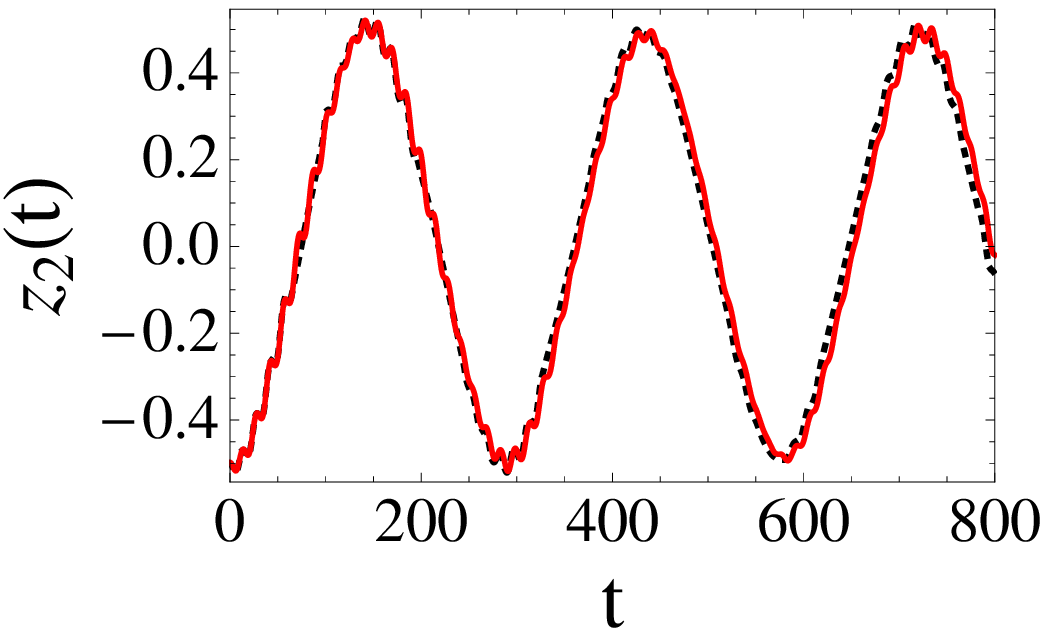}}
\centerline{\includegraphics[width=4cm,clip]{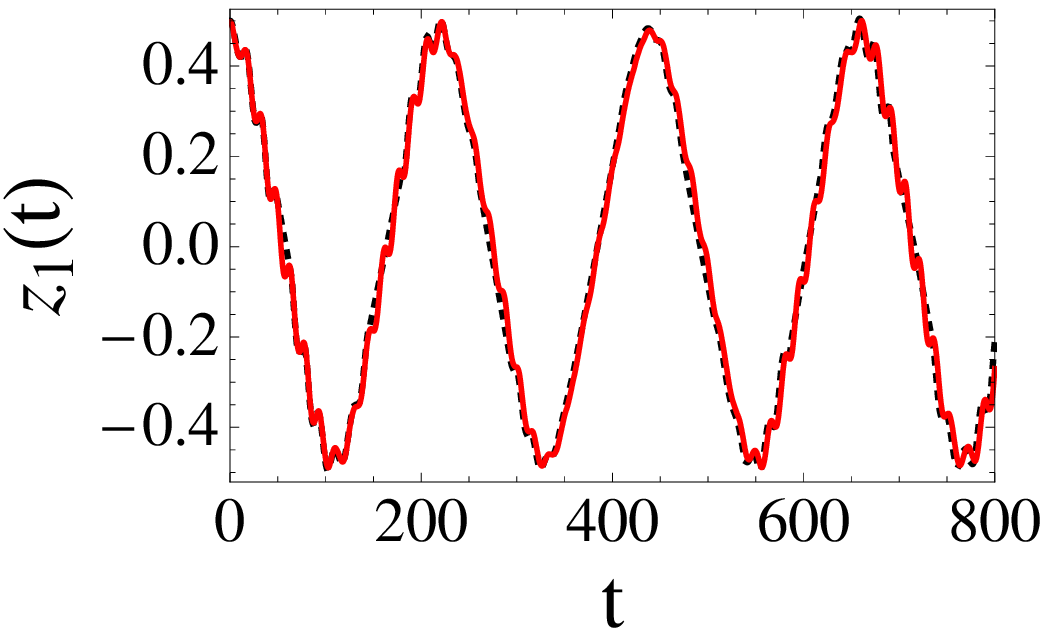}
\includegraphics[width=4cm,clip]{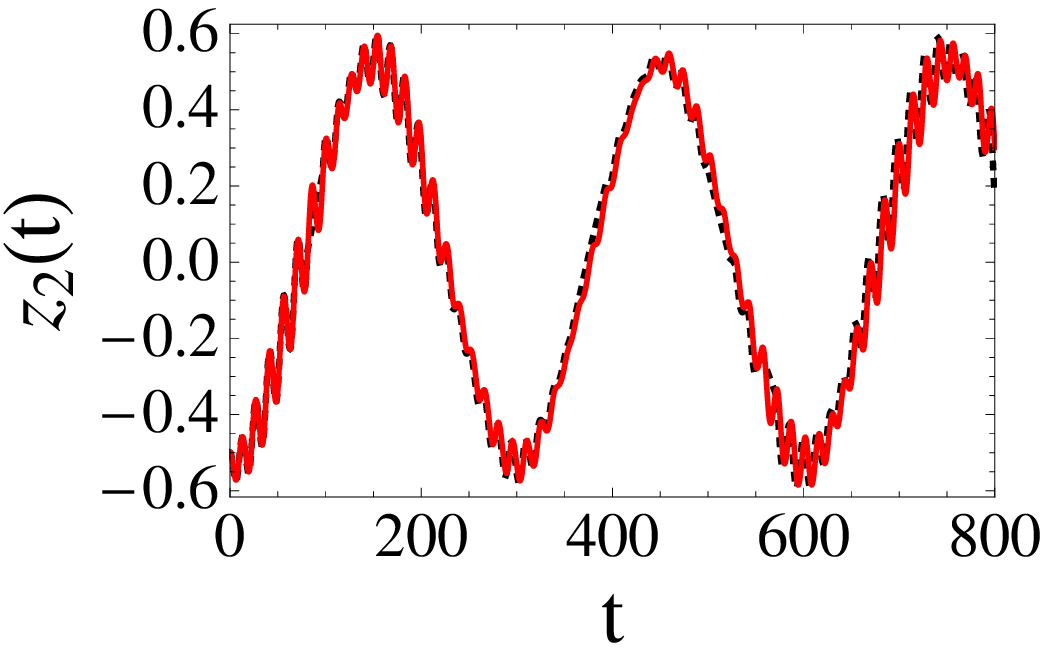}}
\centerline{\includegraphics[width=4cm,clip]{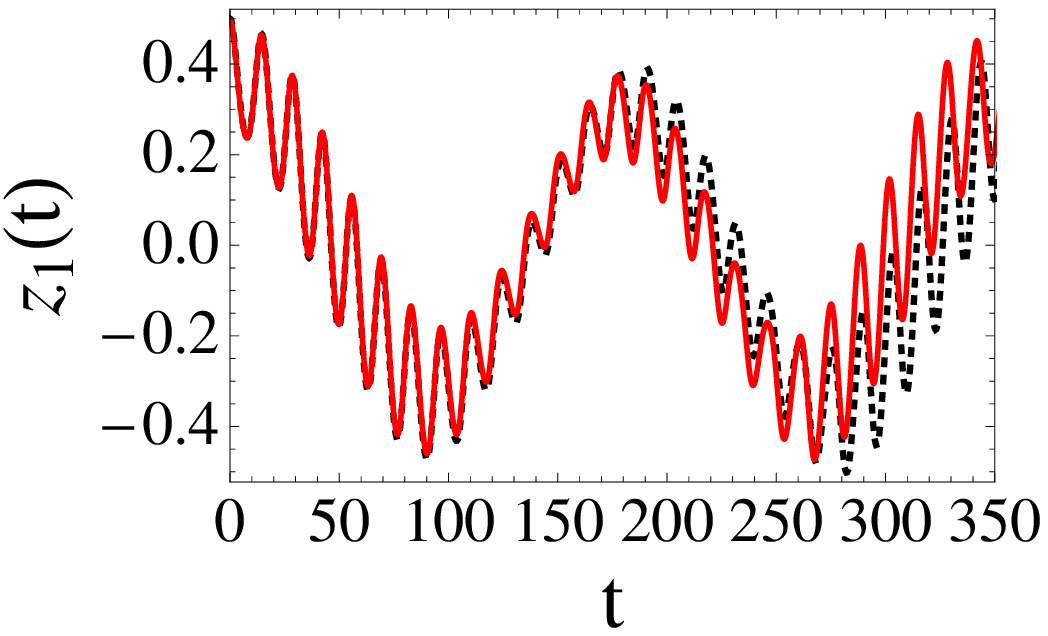}
\includegraphics[width=4cm,clip]{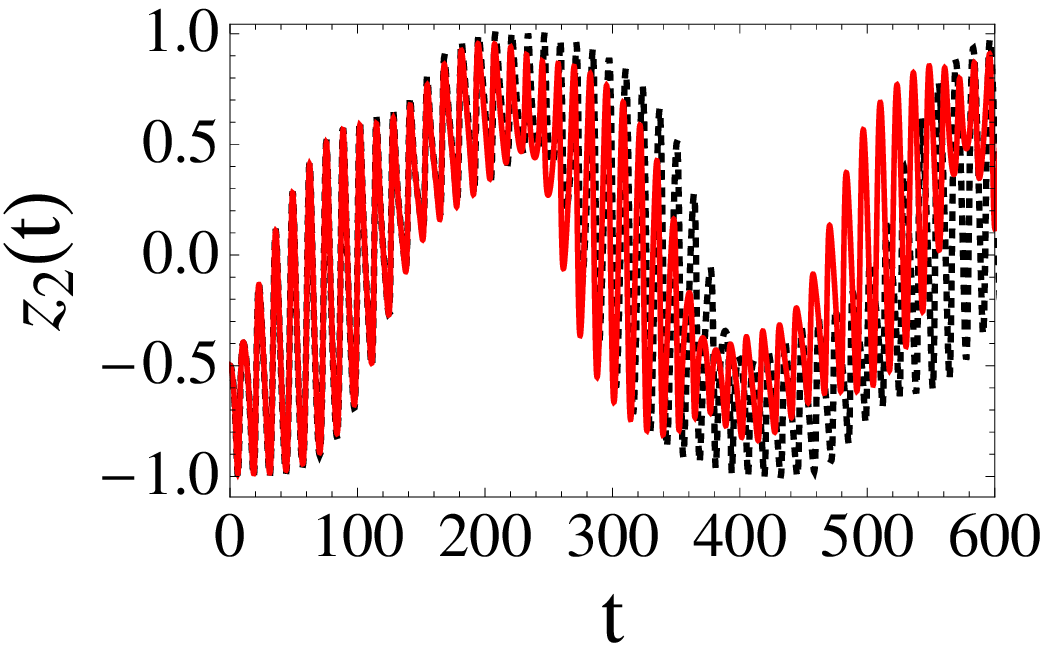}}
\caption{Fractional imbalances $z_{1}$ and $z_{2}$ of the two components vs
time $t$ in the Josephson regime.
Solid lines: 1D GPE, Eqs. ({\protect\ref{1dGPE}}). Dotted lines:
finite-mode equations ({\protect\ref{odesztheta}})
-({\protect\ref{nitetaalpha}}).
Upper panels: $\tilde{\Omega}=0$ (no Rabi coupling);
central panels: $\tilde{\Omega}=K$; third row: $\tilde{\Omega}=3K$;
lower panels: $\tilde{\Omega}=20K$. The  parameters of the
double-well potential (\ref{dwp}) are ${\cal A}=1$ and $z_0=3$.
$K_{1}=K_{2}\equiv K=4.955\times 10^{-3}$, $U_{1}=U_{2}\equiv U=0.1K$, and
$U_{12}=-0.01U$. Other parameters:
$K_{c}=-3.684\times 10^{-6}$, $V=2.268\times 10^{-7}$, $K_{c,12}=-0.01K_c$,
$V_{12}=-0.005 V$; $E_{2}-E_{1}=-0.39$, as in Ref.
\protect\cite{cornell}. Initial conditions: $N_1(0)=200$, $N_2(0)=100$
$z_{1}(0)=0.5=-z_{2}(0)$, $\protect\theta_{n}(0)=0$,
$\protect\gamma_{\alpha}(0)=0$.
Time in units of $\protect\omega_{\bot }^{-1}$,
lengths in units of $a_{\bot}$, and energies in units of
$\hbar \protect\omega _{\bot}$.}
\label{fig1}
\end{figure}

\begin{figure}[ht]
\centerline{\includegraphics[width=4.2cm,height=3.5cm,clip]{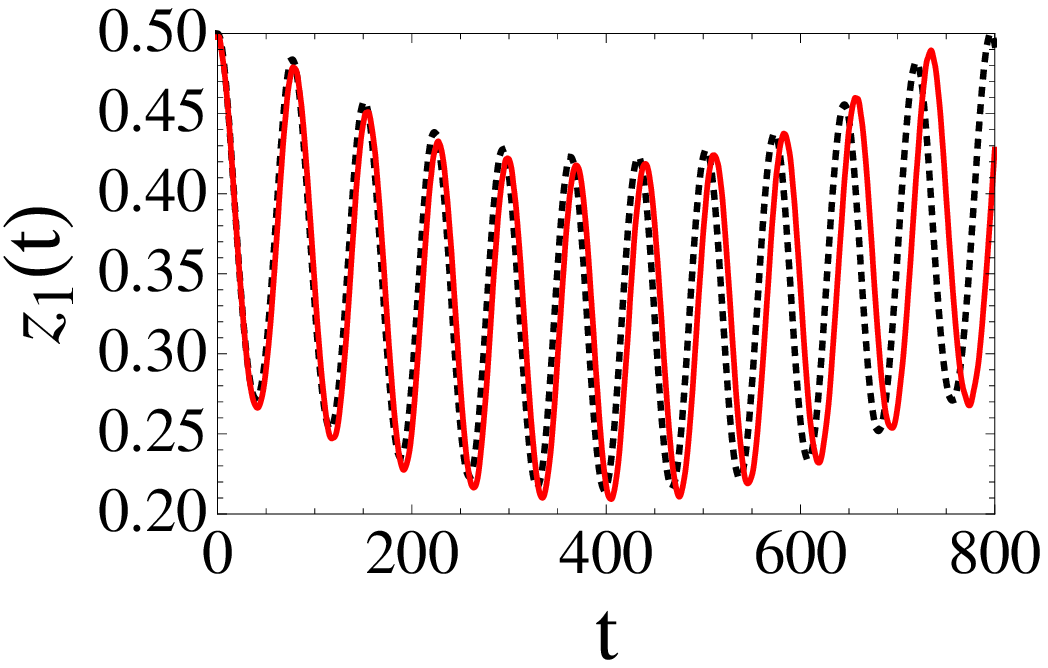}
\includegraphics[width=4.2cm,height=3.5cm,clip]{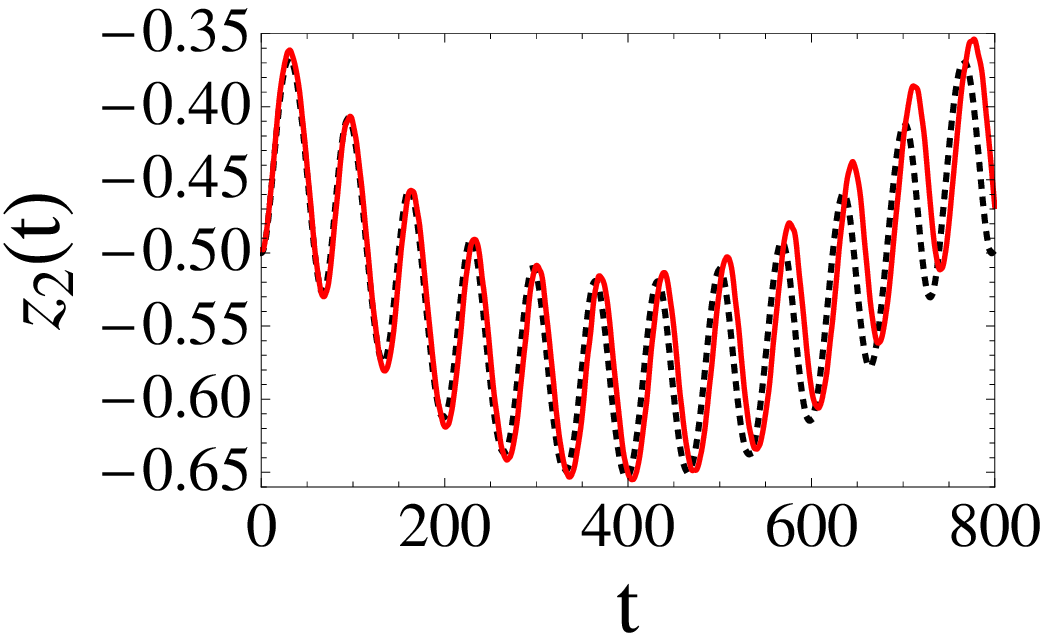}}
\centerline{\includegraphics[width=4.2cm,,height=3.5cm,clip]{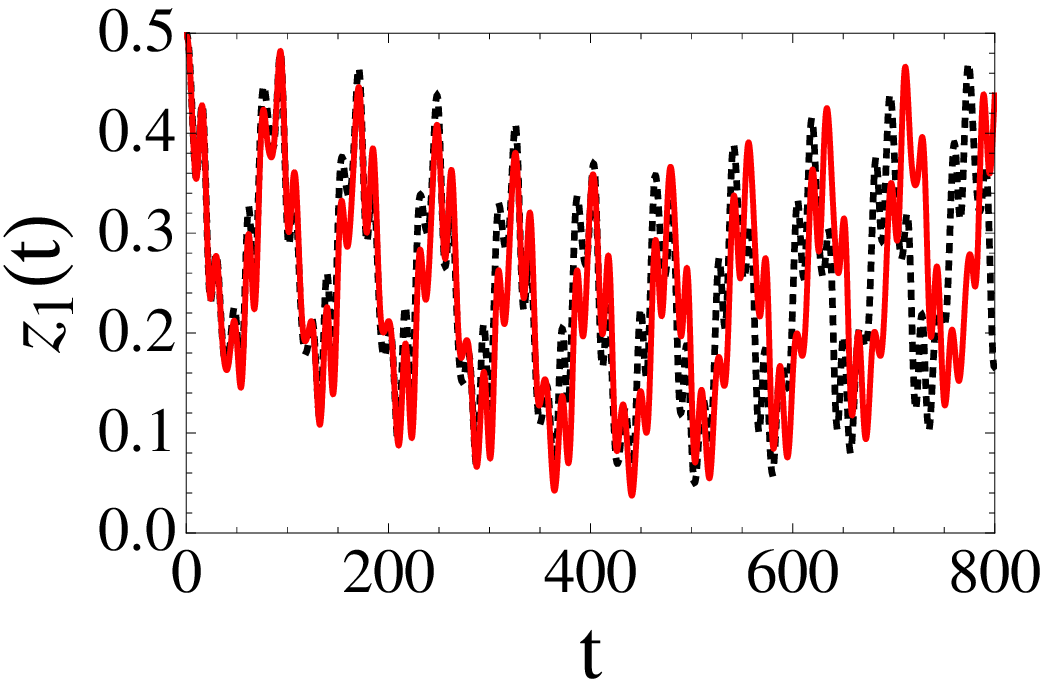}
\includegraphics[width=4.2cm,,height=3.5cm,clip]{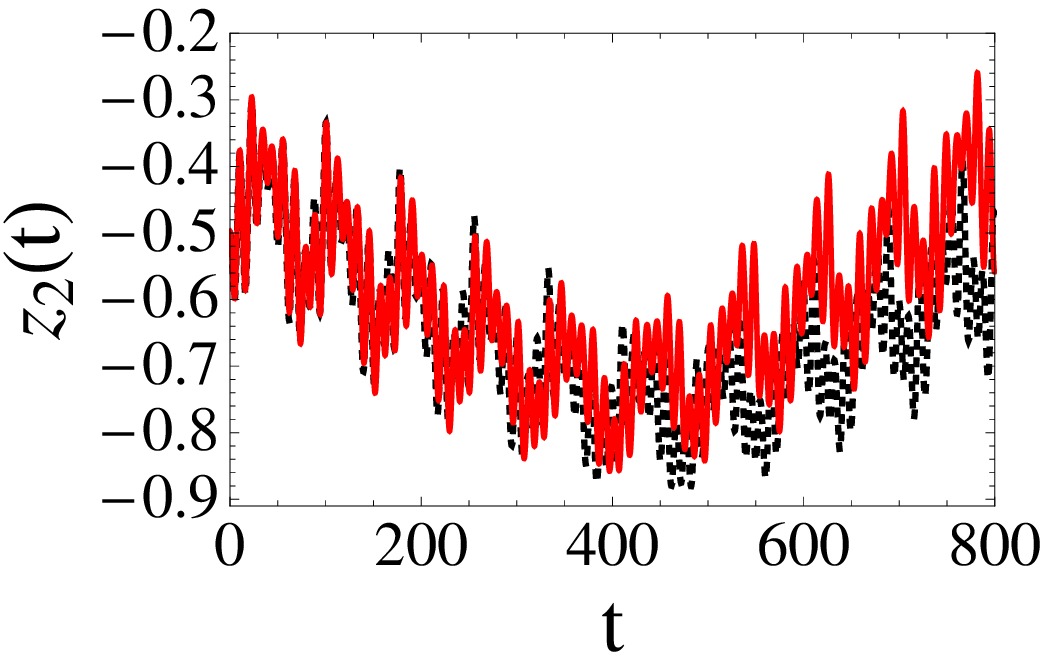}}
\centerline{\includegraphics[width=4.2cm,height=3.5cm,clip]{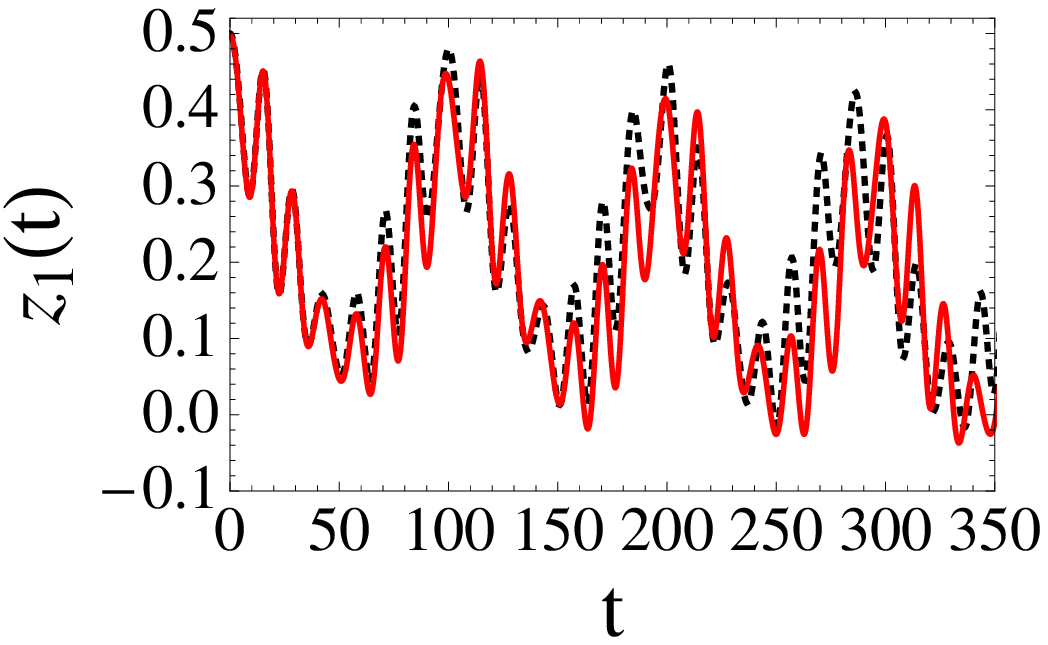}
\includegraphics[width=4.2cm,height=3.5cm,clip]{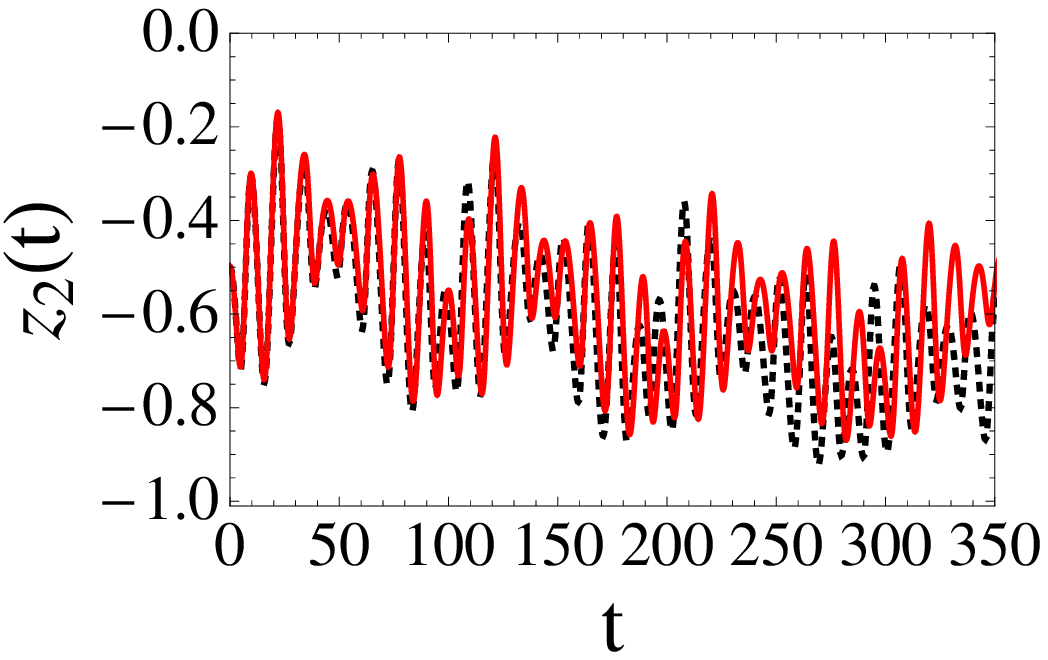}}
\caption{Fractional imbalances $z_{1}$ and $z_{2}$ of the two components vs
time $t$ in the self-trapping regime.
Solid lines: 1D GPE, Eqs. ({\protect\ref{1dGPE}}). Dotted lines:
finite-mode equations ({\protect\ref{odesztheta}})
-({\protect\ref{nitetaalpha}}).
Upper panels: $\tilde{\Omega}=0$ (no Rabi coupling);
central panels: $\tilde{\Omega}=5K$; lower panels: $\tilde{\Omega}=10K$.
We choose $K_{1}=K_{2}\equiv K=4.955\times 10^{-3}$,
$U_{1}=U_{2}\equiv U=0.1K$, and $U_{12}=-2U$, $K_{c,12}=-2K_c$,
$V_{12}=-V$. Other parameters, initial conditions
and units as in Fig. \ref{fig1}.}
\label{fig2}
\end{figure}

To verify the reliability of the finite-mode approximation which leads to the
ODEs (\ref{odesztheta}) and (\ref{nitetaalpha}))-(\ref{nitetaalpha}),
we compare the evolution of fractional imbalances
$z_{n}$, as predicted by this system, with results of direct numerical
simulations of the 1D GPEs (\ref{1dGPE}) both when the fractional imbalances
$z_{n}(t)$ oscillate around a zero-time averaged value and when the
time-averaged value of $\langle z_{n}(t) \rangle \neq 0$, that is the
self-trapping regime. For the oscillations characterized by $\langle z_{n}(t)
\rangle = 0$, the results of the comparison are presented in Fig. \ref{fig1}.
This figure shows a good agreement, especially when the Rabi
coupling, $\tilde{\Omega}$, is small enough. At larger values of $\tilde{%
\Omega}$, the finite-mode approximation demonstrates a deviation, which
accumulates at sufficiently long times.

In Fig. \ref{fig2}, we report the above
comparison between GPEs and ODEs
when the fractional imbalances are both self-trapped. Also in this
case the distance between the predictions from the two approaches increases for
long times when $\tilde \Omega$ is big enough.
It is worth to observe that the authors of Ref. \cite{diaz2} have performed
the comparison between the predictions of GPEs and those ones ODEs deriving
from the two-mode approximation only when the fractional
imbalances oscillate around zero.
Julia-Diaz and co-workers, moreover, have integrated the aforementioned ODEs
under the assumptions of small imbalances, and small intra- and inter-species
phase differences \cite{diaz2}.

\begin{figure}[h!]
\centerline{\includegraphics[width=4.cm,clip]{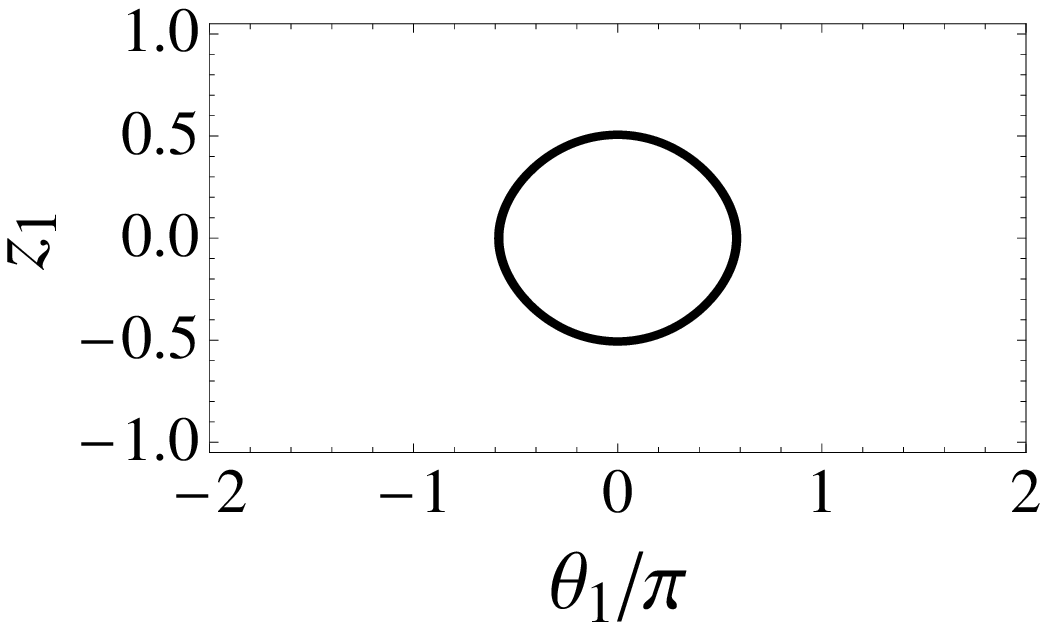}
\includegraphics[width=4.cm,clip]{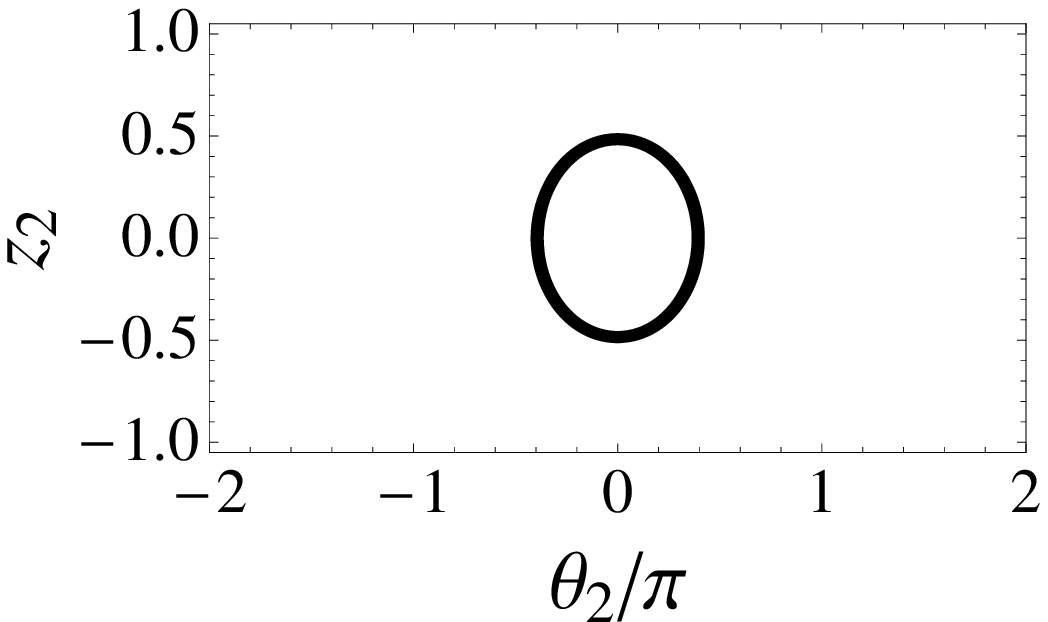}
} \centerline{\includegraphics[width=4cm,clip]{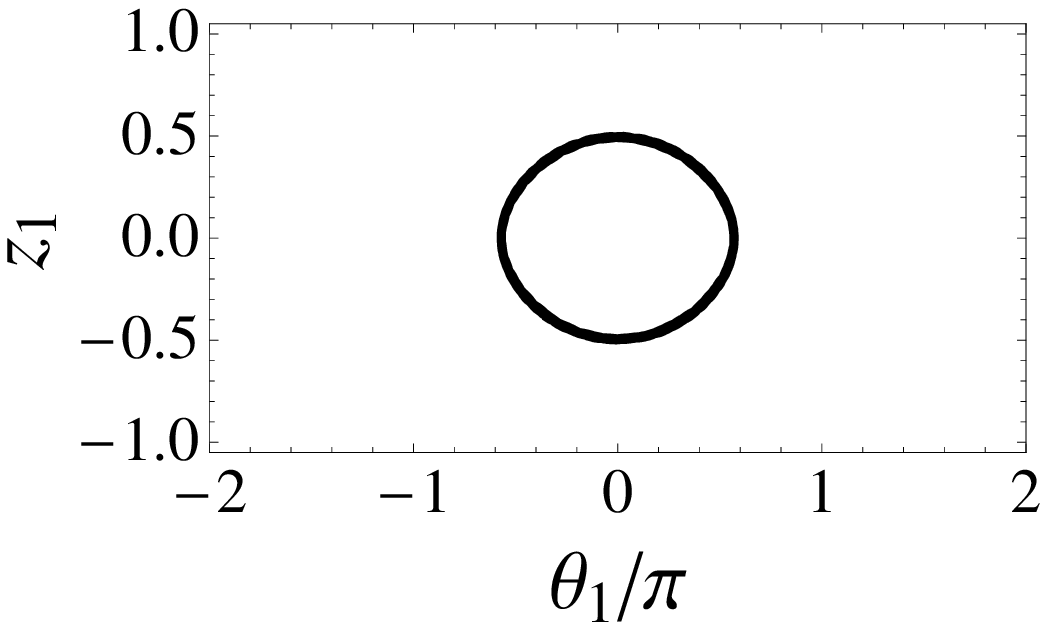}
\includegraphics[width=4cm,clip]{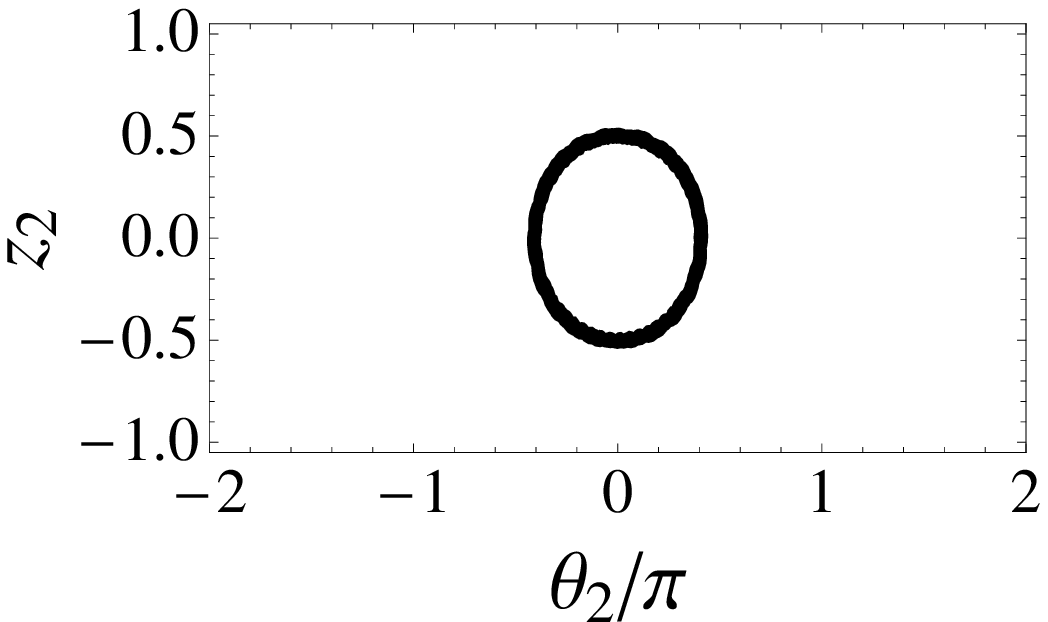}
} \centerline{\includegraphics[width=4cm,clip]{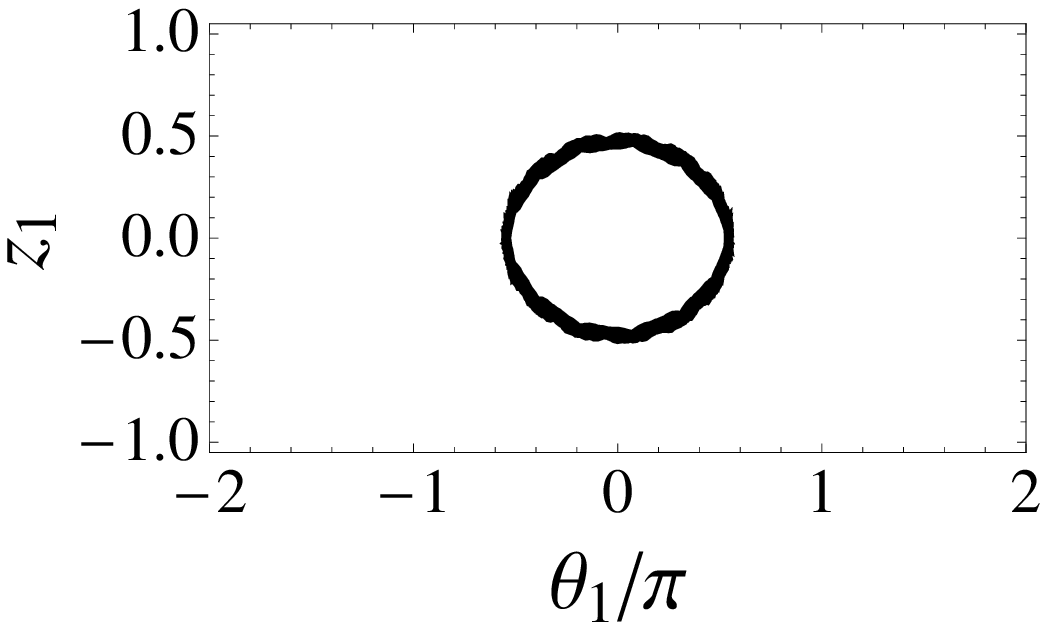}
\includegraphics[width=4cm,clip]{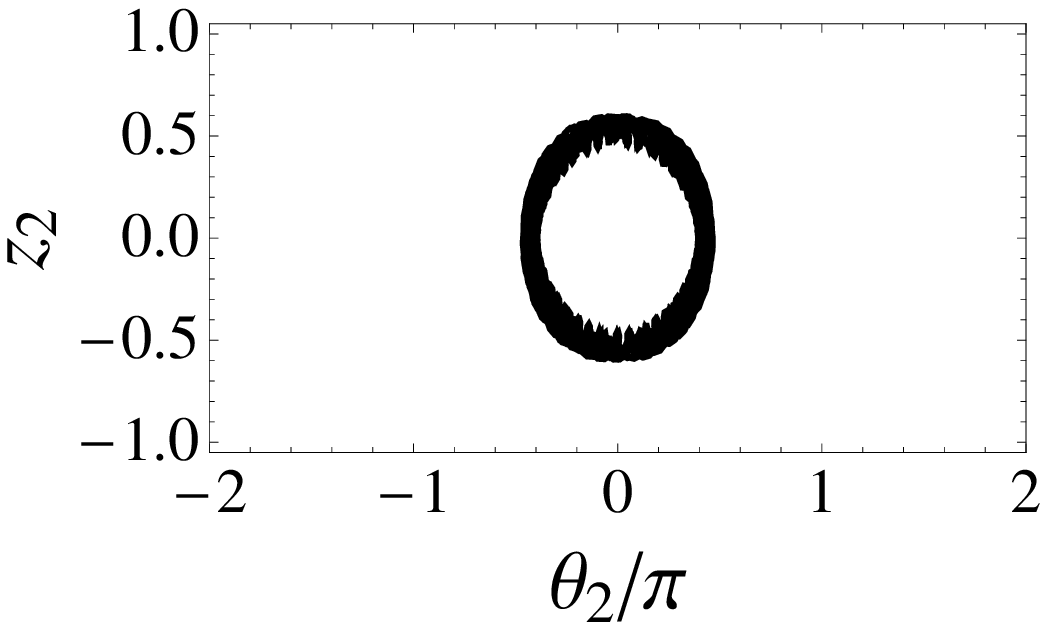}
} \centerline{\includegraphics[width=4cm,clip]{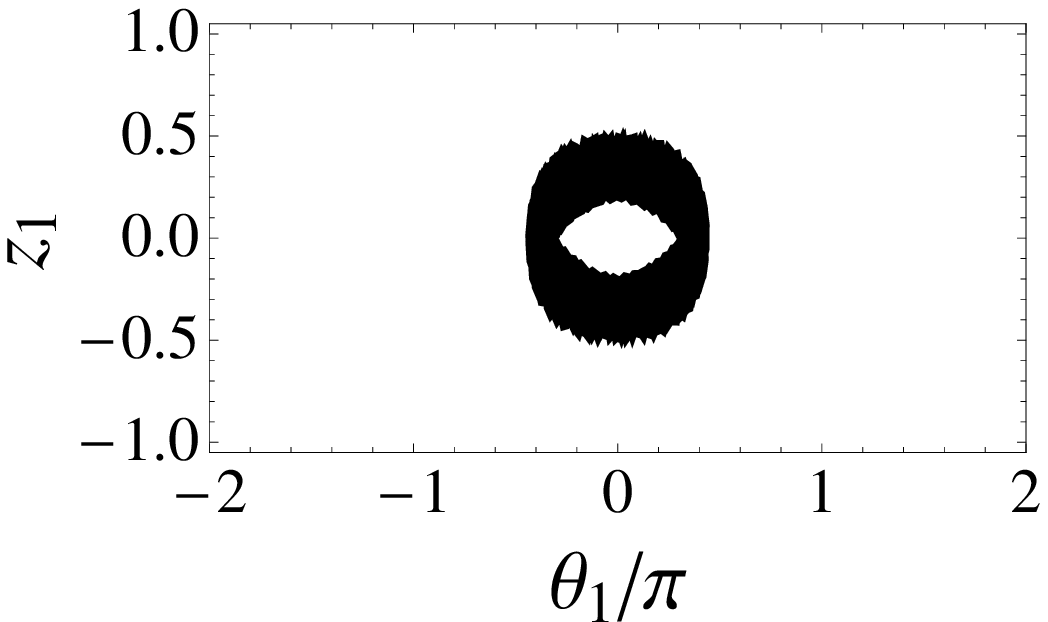}
\includegraphics[width=4cm,clip]{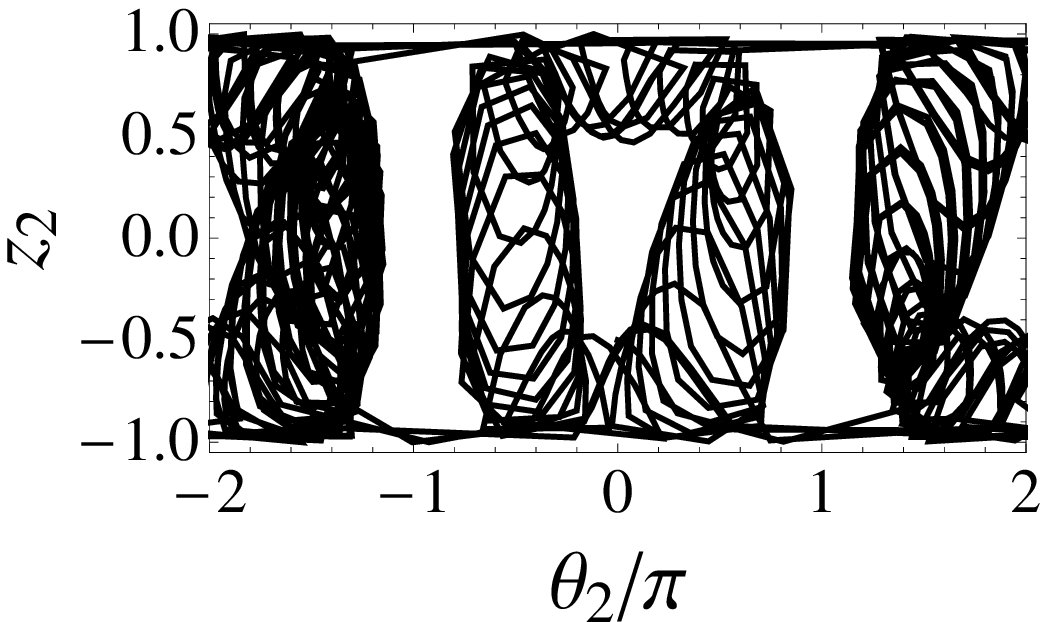}}
\caption{Josephson regime. The dynamics of the finite-dimensional system in the
planes of\ fractional imbalances $z_{n}(t)$ and phases
$\protect\theta _{n}(t)$. The parameters, initial conditions, and
units as in Fig. \ref{fig1}.}
\label{fig3}
\end{figure}

\begin{figure}[h]
\centering
\epsfig{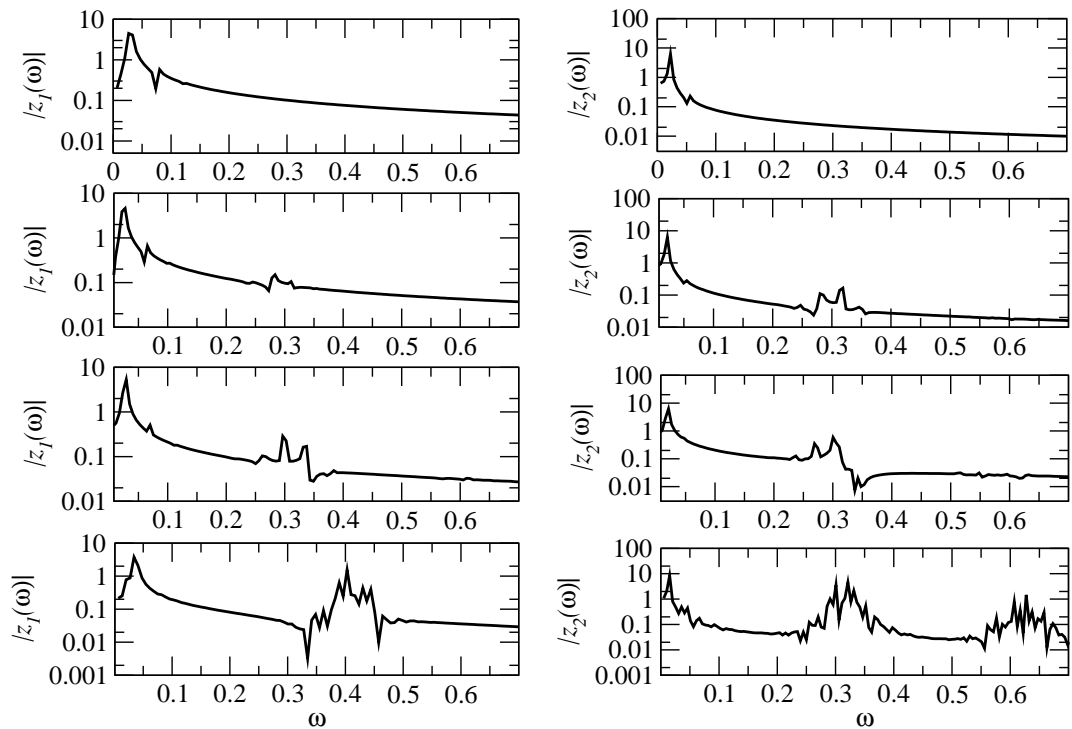}
\caption{Josephson regime. The absolute value of $|z_{n}(\protect\omega )|$
of the Fourier transform of the fractional
imbalances vs. frequency $\protect\omega $. Note that the vertical axis is in
logarithmic scale. The parameters, initial
conditions, and units as in Fig. \ref{fig1}.}
\label{fig4}
\end{figure}

%\begin{figure}[h!]
%\centerline{\includegraphics[width=4.cm,clip]{ft1.eps}
%\includegraphics[width=4.cm,clip]{ft2.eps}
%} \centerline{\includegraphics[width=4cm,clip]{ft3.eps}
%\includegraphics[width=4cm,clip]{ft4.eps}
%} \centerline{\includegraphics[width=4cm,clip]{ft5.eps}
%\includegraphics[width=4cm,clip]{ft6.eps}
%} \centerline{\includegraphics[width=4cm,clip]{ft7.eps}
%\includegraphics[width=4cm,clip]{ft8.eps}}

\begin{figure}[h!]
\centerline{\includegraphics[width=4.2cm,height=3.5cm,clip]{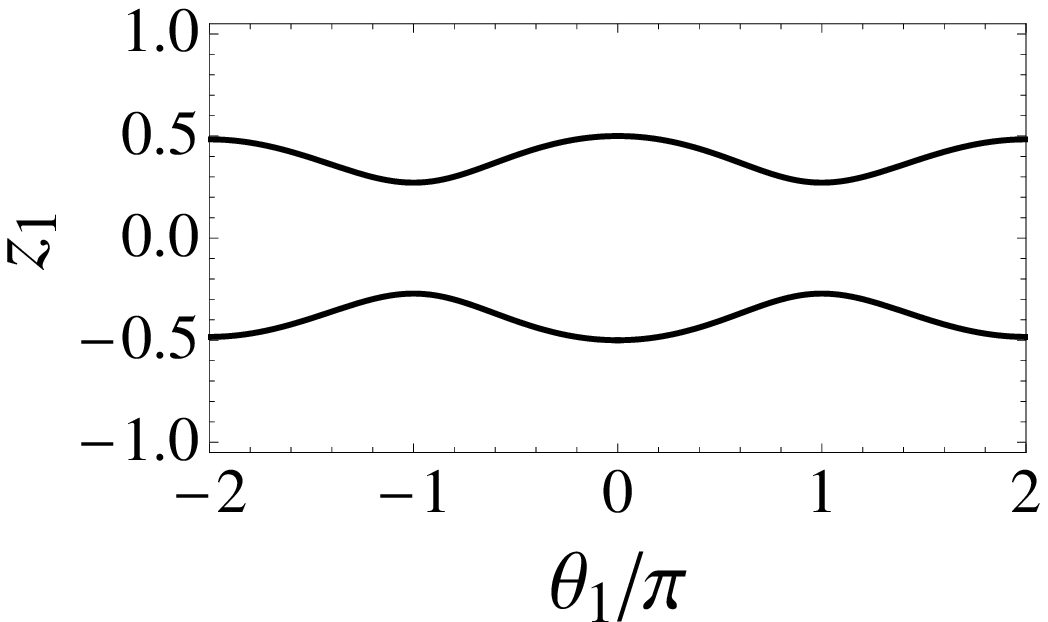}
\includegraphics[width=4.2cm,height=3.5cm,clip]{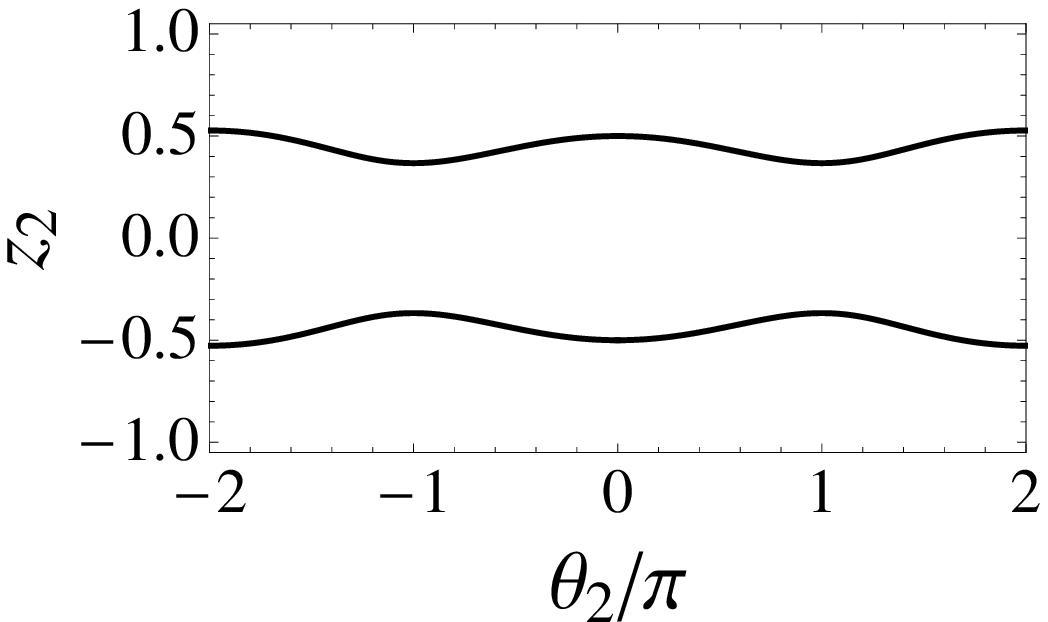}}
\centerline{\includegraphics[width=4.2cm,height=3.5cm,clip]{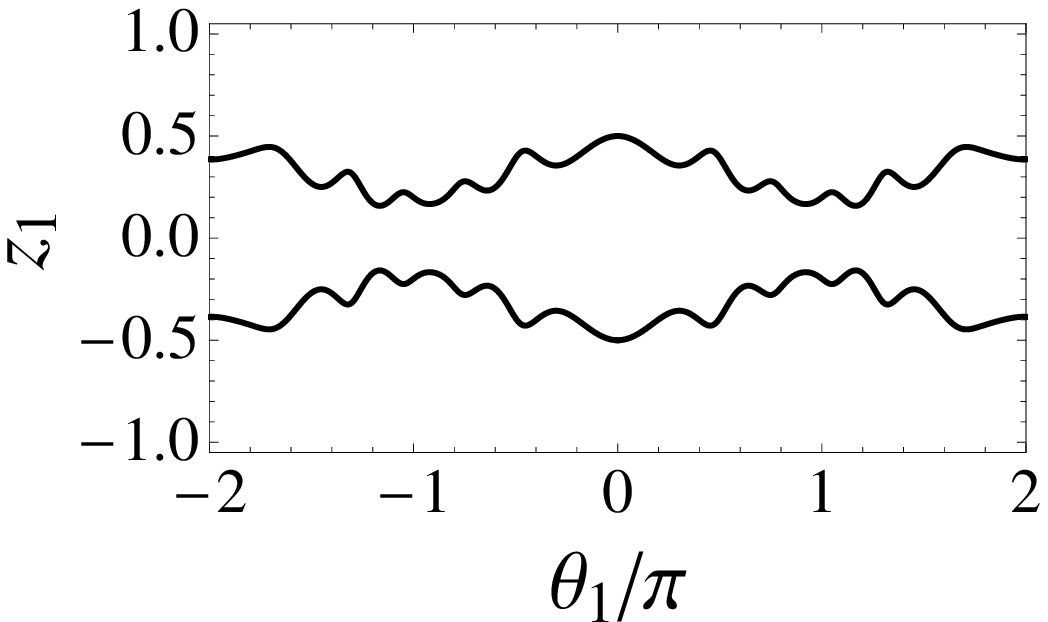}
\includegraphics[width=4.2cm,height=3.5cm,clip]{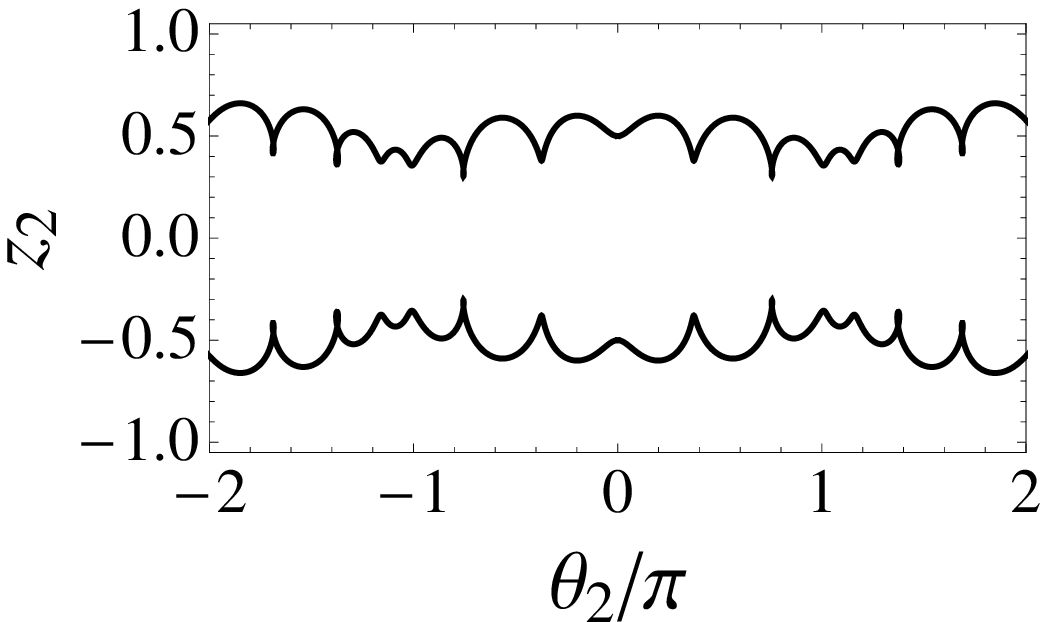}}
\centerline{\includegraphics[width=4.2cm,height=3.5cm,clip]{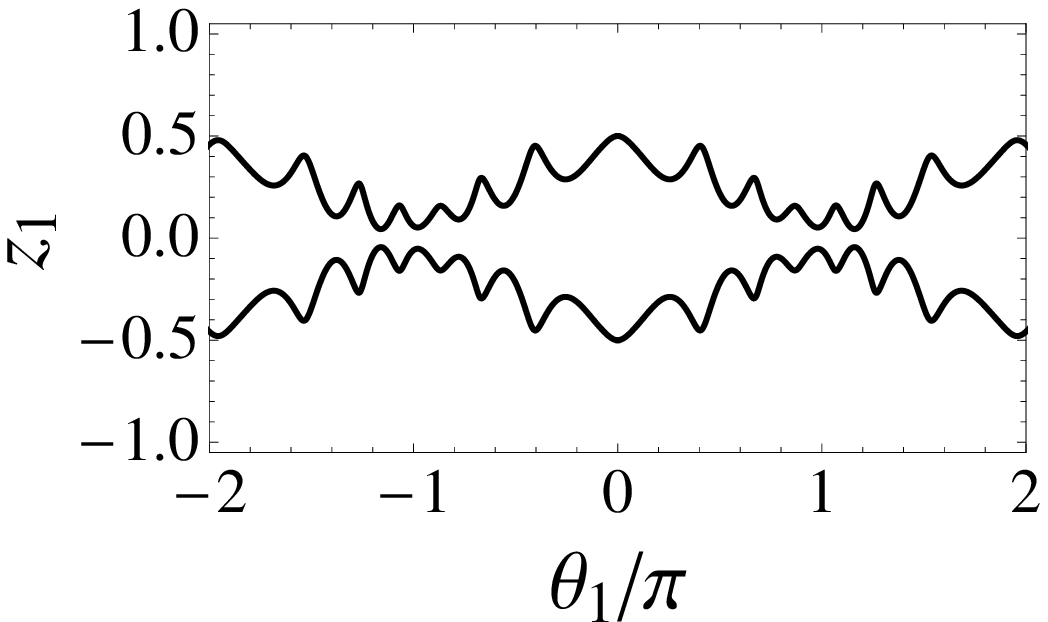}
\includegraphics[width=4.2cm,height=3.5cm,clip]{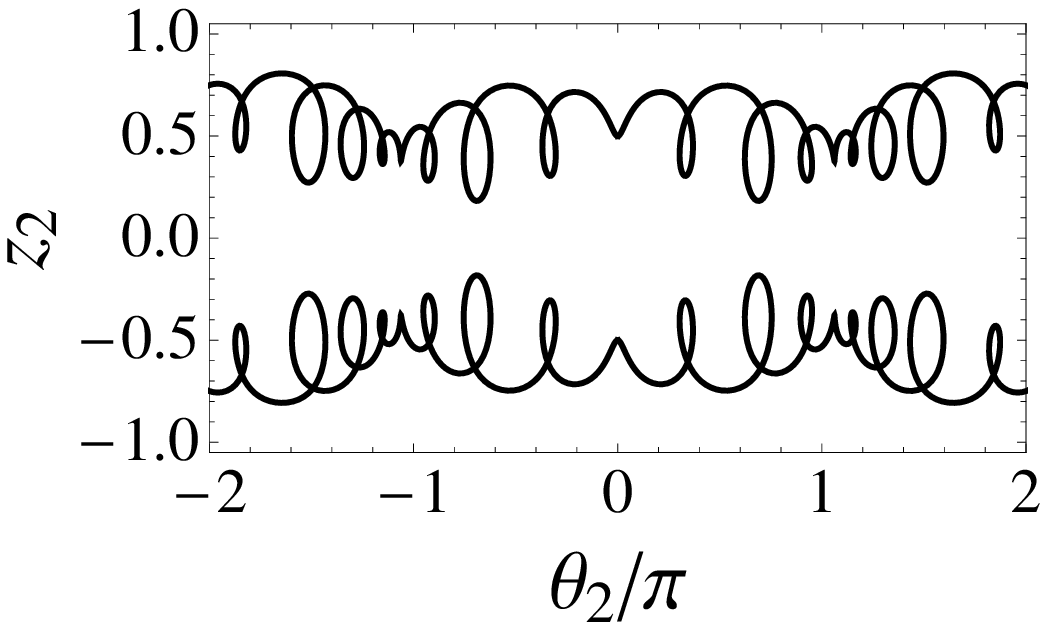}}
\caption{Self-trapping regime.
The dynamics of the finite-dimensional system in the
planes of\ fractional imbalances $z_{n}(t)$ and phases
$\protect\theta _{n}(t)$. The parameters, initial conditions, and
units as in Fig. \ref{fig2}.}
\label{fig5}
\end{figure}

\begin{figure}[h]
\centering
\epsfig{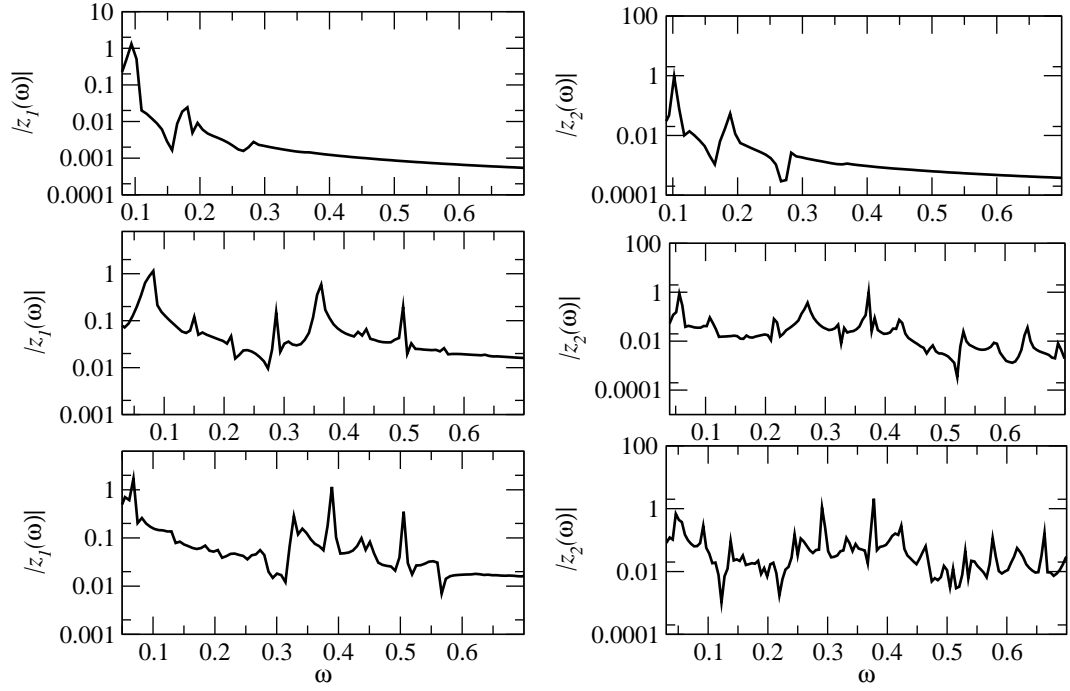}
\caption{Self-trapping regime. The absolute value of $|z_{n}(%
\protect\omega )|$ of the Fourier transform of the fractional
imbalances vs. frequency $\protect\omega $ in the self-trapping regime.
Note that the vertical axis is in
logarithmic scale. The parameters, initial conditions, and units as in Fig. \ref{fig2}.}
\label{fig6}
\end{figure}

%\begin{figure}[h!]
%\centerline{\includegraphics[width=4.2cm,height=3.5cm,clip]{ftst1.eps}
%\includegraphics[width=4.2cm,height=3.5cm,clip]{ftst2.eps}
%} \centerline{\includegraphics[width=4.2cm,height=3.5cm,clip]{ftst3.eps}
%\includegraphics[width=4.2cm,height=3.5cm,clip]{ftst4.eps}
%} \centerline{\includegraphics[width=4.2cm,height=3.5cm,clip]{ftst5.eps}
%\includegraphics[width=4.2cm,height=3.5cm,clip]{ftst6.eps}}

\section{Josephson and Rabi oscillations mixing and the self-trapped
dynamics}

One of the central theme of this work is the interplay of the Josephson oscillations
and Rabi oscillations. We start with the situation in which only the Josephson
coupling is present, i.e., $\tilde{\Omega}=0$, see the upper panels
of Fig. \ref{fig1} and of Figs. \ref{fig3}, \ref{fig4}.
In the other panels of these figures we report the behavior
of the system in the presence of finite values of $\tilde{\Omega}$. One can
see that, the greater is the Rabi coupling between the two species
(hyperfine atomic states), the greater is the dynamical complexity exhibited
by the system, as observed, especially, in the lower panels of Figs.
\ref{fig1}, \ref{fig3} and \ref{fig4}.
Moreover, as shown in the lower panels of Fig. \ref{fig1},
when $\tilde \Omega=20K$, the behaviors of $z_1(t)$ and $z_2(t)$ are
strongly asymmetric with respect to each other; this asymmetry is absent
when $E_2-E_1=0$.

In Fig. \ref{fig3} we plot the dynamics of the finite-dimensional system
in the planes of\ fractional imbalances $z_{n}(t)$ and phases
$\protect\theta _{n}(t)$, using the trapping and input parameters
of Fig. \ref{fig1}. The figure shows the motion is fully periodic
for $\tilde{\Omega}=0$, than it becomes quasi-periodic
and finally aperiodic by increasing the value of $\tilde{\Omega}$.
To quantify the increasing of complexity with the growth of
$\tilde{\Omega}$, in Fig. \ref{fig4} we show
the power spectrum of the oscillations,
represented by the absolute value, $|z_{n}(\omega )|$, of the Fourier
transform of $z_{n}(t)$, %{\color{red}
denoting by $\bar \omega_{1}$ the frequency associated to the
maximum of $|z_{1}(\omega )|$, and by $\bar \omega_{2}$ the one
associated to the maximum of $|z_{2}(\omega )|$. The frequencies
$\bar \omega_{n}$ are the fundamental frequencies of  $z_n(t)$, i.e.
the frequencies of the carrier waves of $z_n(t)$.
 %{\color{red}} 
Let us focus on Table 1 obtained with $N_1(0)=200>N_2(0)=100$. From the
data there reported, we see that - whatever is the value of the Rabi
coupling $\tilde \Omega$ - $\bar \omega_{1} > \bar \omega_{2}$. Then
the multi-particle tunneling period associated to $\bar \omega_{1}$,
$\bar T_{1}$, given by $2\pi/\bar \omega_{1}$, is smaller than the
one associated to $\bar \omega_{2}$, $\bar T_{2}$, given by
$2\pi/\bar \omega_{2}$.  This is due to the fact that - within a
mechanical analogy - the ODEs equations
(\ref{odesztheta})-(\ref{nitetaalpha}) describes two coupled
pendula.  At least in the presence of sufficiently weak
inter-species interactions, the mass of each pendulum is related to
the inverse of the particles number $N_n$ of the $n$th species, as
discussed for a single component in Ref. \cite{smerzi}. From the
Table 1, it is possible infer as well that the greater is $\tilde
\Omega$ the greater is the ratio $\bar T_{2}/\bar T_{1}$. On the
other hand, if we keep fixed the Rabi coupling and increase $N_1(0)$
with respect to $N_2(0)$, the period $\bar T_{1}$ will be smaller
and smaller if compared with $\bar T_{2}$.
%Finally, from
%Table 1 again, we can see that the greater is $\tilde \Omega$
%the greater (the smaller) is $\bar \omega_{1}$
%($\bar \omega_{2}$).
Moreover,  from Table 1 again, we see that the relative changes $r_n$ of $\bar
\omega_{n}$ with respect to their values at $\tilde \Omega=0$ increase by increasing $\tilde \Omega$.%}

%{\color{red}
From Fig. \ref{fig4} it is possible to gain physical insight in the
dynamics of the system, especially for large values of the Rabi coupling $\tilde
\Omega$. We can see that when $\tilde \Omega$ is sufficiently large, $|z_{n}(\omega )|$ exhibits a multi-peak structure related to the
appearance of frequencies different from the fundamental one. This reflects in
an increasing number of harmonics involved in $z_n(t)$ (see, in particular, the lower panels of Fig.
\ref{fig1} where $\tilde \Omega=20 K$) and, accordingly, in an increasing degree
of complexity. By analyzing Fig. \ref{fig4} again, one can
conclude that the power spectrum approaches that of random oscillations as
$\tilde{\Omega}$ increases. This, immediately, reflects on a quite complicated
dynamics when we focus on the plane $z_n(t)-\theta_n(t)$ (see,
in particular, the lower panels in Fig. \ref{fig3} where $\tilde \Omega=20 K$).
Finally, note that the complexity pertaining to high values of $\tilde \Omega$ increases if the dynamics is
observed on sufficiently long time scales as shown in the lower panels of Fig. \ref{fig3} and Fig. \ref{fig4}.
%}

\begin{center}
\begin{tabular}{|c|c|c|c|c|}
\hline
~~~$\tilde \Omega$~~~ & ~~~$\bar \omega_{1}$~~~& ~~~$\bar
\omega_{2}$~~~ &~~~$r_1$~~~ & ~~~$r_2$~~~\\
\hline
 $0$ & $0.0268$ & $0.0224$  & $0$ & $0$  \\
 $K$ & $0.0272$ & $0.0220$ & $0.015$ & $0.018$ \\
 $3K$ & $0.0291$ & $0.0211$ & $0.086$ & $0.060$\\
 $20K$ & $0.0341$ & $0.0153$ & $0.27$ & $0.32$ \\
 \hline
\end{tabular}
\end{center}
Table 1. {\small Spectra of Fig. \ref{fig4}.
Frequencies $\bar \omega_{n}$ of the maxima of
$|z_{n}(\omega )|$ for different values of $\tilde{\Omega}$ and
their relative changes $r_{n}$ with
respect to the absence of the Rabi coupling.}

\vskip 0.5cm

For the investigation of the self-trapped dynamics
let us consider again Fig. \ref{fig2}
and also Figs. \ref{fig5} and \ref{fig6}.
In Fig. \ref{fig5} we report the dynamics in the planes
$\left(z_{n},\theta \right)$,
while in Fig. \ref{fig6}) we show the absolute value
$|z_{n}(\omega)|$ of the Fourier
transform of $z_{n}(t)$. From these figures, we see that in
correspondence to high value of the linear coupling constant, an increasing
complexity is observed within the dynamics supported by the junction.

\begin{figure}[h]
\centering
\epsfig{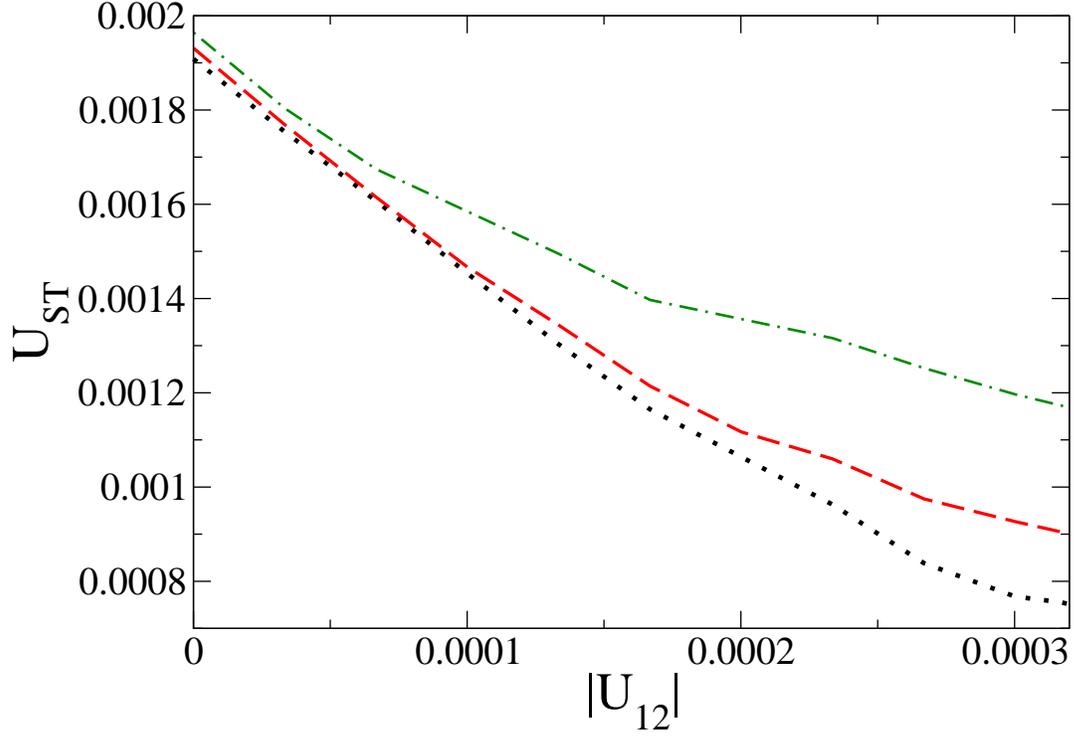}
\caption{Self-trapping crossover value $U_{ST}$ as a function of the
inter-species interaction $|U_{12}|$ for
different values of $\tilde{\Omega}$.
Dotted line: $\tilde{\Omega}=0$; dashed line:
$\tilde{\Omega}=5 K$; dot-dashed line: $\tilde{\Omega}=10 K$.
The double-well potential, initial conditions and units
as in Fig. \ref{fig2}.}
\label{fig7}
\end{figure}

It is important to stress that it exists a crossover
value of the intra-species
interaction, say $U_{ST}$, (note that we are setting $U_{1}=U_{2}
\equiv U$), such that the two components are both self-trapped
when $U>U_{ST}$. By integrating the ODEs ({\protect\ref%
{odesztheta}})-({\protect\ref{nitetaalpha}}), we study $U_{ST}$
as a function of the absolute value of the
inter-species interaction amplitude $U_{12}$ for different values of $\tilde
\Omega$. The results of this study are presented in Fig. \ref{fig7}.
>From the plots there reported, we can see
that, once fixed $U_{12}$, the greater is $\tilde \Omega$, the greater is
the value of the intra-species interaction necessary to achieve the
self-trapping.

\section{Conclusion}

In this work, we have introduced the model which allows one to study the
interplay of the Josephson and Rabi oscillations in a binary Bose-Einstein
condensate trapped in the double-well potential structure.
The Rabi coupling is provided by the interconversion
between the two species of the condensate, which represent distinct
hyperfine states of the same atom.

To capture core features of the dynamics, we have derived a finite-mode
approximation with four degrees of freedom, that represent the populations
of the two species in the two symmetric potential wells. Comparison to full
simulations of the underlying Gross-Pitaevskii system demonstrates
that the truncated system provided for a reasonable accuracy. Systematic simulations of the
system reveal the transition from regular Josephson oscillations to complex
dynamics with the increase of the Rabi interconversion rate.
%{\color{red}
Within the framework of this analysis, we have discussed the
possibility to infer, at least at a qualitative level, the behavior
of the fractional imbalances and the dynamics in the plane
$z_n(t)-\theta_n(t)$ proceeding from the Fourier analysis of the
fractional imbalances, especially
for large Rabi couplings.%}

We have analyzed the dynamics of the atomic Josephson junction when both the
components are self-trapped as well. We have shown that also in this case the
truncated system gives rise to reliable predictions since the good agreement
with the predictions of the associated Gross-Pitaevskii system. We have studied
the influence of the linear coupling on the self-trapping onset.

\section*{Appendix}

In this appendix, we discuss the path followed in deriving the evolution
equations for the fractional imbalances $z_n=(N_{n}^{L}-N_{n}^{R})/N_n$
and the
intra-species relative phases $\theta_n=\theta_{n}^{R}-\theta_{n}^{L}$, i.e.
Eq. (\ref{odesztheta}) and the Eq. (\ref{nitetaalpha}).
We start by deriving the effective Lagrangian $L_{\mathrm{eff}}$
in terms of variables $N_{n}^{\alpha }$ and $\theta_{n}^{\alpha }$ :
\begin{eqnarray}
L_{\mathrm{eff}} &=&\sum_{n=1,2}\bigg[-\hbar \dot{\theta}_{n}^{L}N_{n}^{L}-
\hbar \dot{\theta}_{n}^{R}N_{n}^{R}-E_{n}^{L}N_{n}^{L}-E_{n}^{R}N_{n}^{R}
\nonumber
\label{lagrangianbis} \\
&+&2K_{n}\sqrt{N_{n}^{L}N_{n}^{R}}\cos (\theta _{n}^{R}-\theta _{n}^{L})
\nonumber \\
&-&\big(\frac{U_{n}^{L}}{2}(N_{n}^{L})^{2}+\frac{U_{n}^{R}}{2}(N_{n}^{R})^{2}
\big)  \nonumber \\
&-&2K_{c,n}N_{n}\sqrt{N_{n}^{L}N_{n}^{R}}\cos (\theta _{n}^{R}-\theta
_{n}^{L})  \nonumber \\
&-&V_{n}N_{n}^{L}N_{n}^{R}(2+\cos 2(\theta _{n}^{R}-\theta _{n}^{L})\bigg]
\nonumber
\\
&-&U_{12}^{L}N_{1}^{L}N_{2}^{L}-U_{12}^{R}N_{1}^{R}N_{2}^{R}  \nonumber \\
&-&2\,K_{c,12}\bigg((N_{1}^{L}+N_{1}^{R})\sqrt{N_{2}^{L}N_{2}^{R}}\,\cos
(\theta _{2}^{R}-\theta _{2}^{L})  \nonumber \\
&+&(N_{2}^{L}+N_{2}^{R})\sqrt{N_{1}^{L}N_{1}^{R}}\,\cos (\theta
_{1}^{R}-\theta _{1}^{L})\bigg)  \nonumber \\
&-&V_{12}\big(N_{1}^{L}N_{2}^{R}+N_{1}^{R}N_{2}^{L}\big)  \nonumber \\
&-&4\,V_{12}\big(\sqrt{N_{1}^{L}N_{1}^{R}}\,\sqrt{N_{2}^{L}N_{2}^{R}}\,\cos
(\theta _{1}^{R}-\theta _{1}^{L})\,\cos (\theta _{2}^{R}-\theta _{2}^{L})%
\big)  \nonumber \\
&-&2\big(R_{12}^{L}\sqrt{N_{1}^{L}N_{2}^{L}}\,\cos (\theta _{2}^{L}-\theta
_{1}^{L})\big)  \nonumber \\
&-&2\big(R_{12}^{R}\sqrt{N_{1}^{R}N_{2}^{R}}\,\cos (\theta _{2}^{R}-\theta
_{1}^{R})\big)  \nonumber \\
&-&2\big(S_{12}^{LR}\sqrt{N_{1}^{L}N_{2}^{R}}\,\cos (\theta _{2}^{R}-\theta
_{1}^{L})\big)  \nonumber \\
&-&2\big(T_{12}^{RL}\sqrt{N_{1}^{R}N_{2}^{L}}\,\cos (\theta _{2}^{L}-\theta
_{1}^{R})\big)\;,
\end{eqnarray}%
where the following constants are introduced:
\begin{eqnarray}
&&E_{n}^{\alpha }=\int dz\,\bigg[\frac{\hbar ^{2}}{2m}\,\left( \frac{d\phi
_{n}^{\alpha }}{dz}\right) ^{2}+  \nonumber  \label{parameters} \\
&&\bigg(\frac{\hbar ^{2}}{2ma_{\bot ,n}^{2}}+\frac{m\omega _{n}^{2}a_{\bot
,n}^{2}}{2}+V_{\mathrm{DWP}}(z)+\frac{(-1)^{n}\,\hbar \,\delta }{2}\bigg)%
\left( \phi _{n}^{\alpha }\right) ^{2}\bigg]  \nonumber \\
&&U_{n}^{\alpha }=\tilde{g}_{n}\int dz\,\left( \phi _{n}^{\alpha }\right)
^{4}  \nonumber \\
&&K_{n}=-\int dz\,\bigg[\frac{\hbar ^{2}}{2m}\frac{d\phi _{n}^{L}}{dz}\frac{%
d\phi _{n}^{R}}{dz}+V_{\mathrm{DWP}}(z)\phi _{n}^{L}\phi _{n}^{R}\bigg]
\nonumber \\
&&K_{c,n}=\tilde{g}_{n}\int dz\,(\phi _{n}^{\alpha
}(z))^{3}\,\phi _{n}^{\beta }(z)  \nonumber \\
&&V_{n}=2\tilde{g}_{n}\int dz\,(\phi _{n}^{\alpha
}(z))^{2}\,(\phi _{n}^{\beta }(z))^{2}  \nonumber \\
&&U_{12}^{\alpha }=\tilde{g}_{12}\int dz\,(\phi _{1}^{\alpha })^{2}(\phi
_{2}^{\alpha })^{2}  \nonumber \\
&&K_{c,12}=\tilde{g}_{12}\int dz\,(\phi _{1}^{\alpha
}(z))^{3}\,\phi _{2}^{\beta }(z)  \nonumber \\
&&V_{12}=\tilde{g}_{12}\int dz\,(\phi _{1}^{\alpha
}(z))^{2}\,(\phi _{2}^{\beta }(z))^{2}  \nonumber \\
&&R_{12}^{\alpha }=\left( \tilde{\Omega}/2\right) \int dz\,\phi _{1}^{\alpha
}(z)\phi _{2}^{\alpha }(z)  \nonumber \\
&&S_{12}^{LR}=\left( \tilde{\Omega}/2\right) \int dz\,\phi _{1}^{L}(z)\phi
_{2}^{R}(z)  \nonumber \\
&&T_{12}^{RL}=\left( \tilde{\Omega}/2\right) \int dz\,\phi _{1}^{R}(z)\phi
_{2}^{L}(z)\;.
\end{eqnarray}
To analyze the finite-mode dynamics induced by Lagrangian (\ref%
{lagrangianbis}), we define the canonical \ momenta conjugate to generalized
coordinates $\hbar \theta _{n}^{\alpha }$:
\begin{equation}
p_{\theta _{n}^{\alpha }}=\frac{1}{\hbar }
\frac{\partial L_{\mathrm{eff}}}{\partial
\dot{\theta}_{n}^{\alpha }}=-N_{n}^{\alpha }\;.  \label{conjugatemoments}
\end{equation}%
Accordingly, the Hamiltonian of the system is written in terms of the
canonical coordinates and momenta as follows:
\begin{eqnarray}
H &=&-\sum_{n=1,2}[p_{\theta _{n}^{L}}E_{n}^{L}+p_{\theta _{n}^{R}}E_{n}^{R}]
\nonumber  \label{hamiltonian} \\
&-&\sum_{n=1,2}2K_{n}^{\alpha }\sqrt{p_{\theta _{n}^{L}}p_{\theta _{n}^{R}}}%
\cos (\theta _{n}^{L}-\theta _{n}^{R})  \nonumber \\
&+&\sum_{n=1,2}\big[\frac{U_{n}^{L}}{2}p_{\theta _{n}^{L}}^{2}+\frac{%
U_{n}^{R}}{2}p_{\theta _{n}^{R}}^{2}  \nonumber \\
&-&2K_{c,n}(p_{\theta _{n}^{R}}+p_{\theta _{n}^{L}})\sqrt{p_{\theta
_{n}^{R}}p_{\theta _{n}^{L}}}\cos (\theta _{n}^{R}-\theta _{n}^{L})
\nonumber \\
&+&V_{n}(2+\cos 2(\theta _{n}^{L}-\theta _{n}^{R}))p_{\theta
_{n}^{L}}p_{\theta _{n}^{R}}\big]  \nonumber \\
&+&U_{12}^{L}p_{\theta _{1}^{L}}p_{\theta _{2}^{L}}+U_{12}^{R}p_{\theta
_{1}^{R}}p_{\theta _{2}^{R}}  \nonumber \\
&-&2\,K_{c,12}\bigg[(p_{\theta _{1}^{L}}+p_{\theta _{1}^{R}})\sqrt{p_{\theta
_{2}^{L}}p_{\theta _{2}^{R}}}\,\cos (\theta _{2}^{R}-\theta _{2}^{L})
\nonumber \\
&+&(p_{\theta _{2}^{L}}+p_{\theta _{2}^{R}})\sqrt{p_{\theta
_{1}^{L}}p_{\theta _{1}^{R}}}\,\cos (\theta _{1}^{R}-\theta _{1}^{L})\bigg]
\nonumber \\
&+&V_{12}\big(p_{\theta _{1}^{L}}p_{\theta _{2}^{R}}+p_{\theta
_{1}^{R}}p_{\theta _{2}^{L}}\big)  \nonumber \\
&+&4\,V_{12}\big(\sqrt{p_{\theta _{1}^{L}}p_{\theta _{1}^{R}}}\,\sqrt{%
p_{\theta _{2}^{L}}p_{\theta _{2}^{R}}}\,\cos (\theta _{1}^{R}-\theta
_{1}^{L})\,\cos (\theta _{2}^{R}-\theta _{2}^{L})\big)  \nonumber \\
&+&2\big(R_{12}^{L}\sqrt{p_{\theta _{1}^{L}}p_{\theta _{2}^{L}}}\,\cos
(\theta _{2}^{L}-\theta _{1}^{L})\big)  \nonumber \\
&+&2\big(R_{12}^{R}\sqrt{p_{\theta _{1}^{R}}p_{\theta _{2}^{R}}}\,\cos
(\theta _{2}^{R}-\theta _{1}^{R})\big)  \nonumber \\
&+&2\big(S_{12}^{LR}\sqrt{p_{\theta _{1}^{L}}p_{\theta _{2}^{R}}}\,\cos
(\theta _{2}^{R}-\theta _{1}^{L})\big)  \nonumber \\
&+&2\big(T_{12}^{RL}\sqrt{p_{\theta _{1}^{R}}p_{\theta _{2}^{L}}}\,\cos
(\theta _{2}^{L}-\theta _{1}^{R})\big)\;.
\end{eqnarray}%

The evolution equations for populations $N_{n}^{\alpha }$ and phases $\theta
_{n}^{\alpha }$ are derived, as the canonical equations, from the
Hamiltonian: (\ref{hamiltonian})
\begin{equation}
\dot{p}_{\theta _{n}^{\alpha }}=-\frac{1}{\hbar }\frac{\partial H}{\partial
\theta _{n}^{\alpha }}\;,\;\dot{\theta}_{n}^{\alpha }=\frac{1}{\hbar }\frac{%
\partial H}{\partial p_{\theta _{n}^{\alpha }}}\;.  \label{canonical}
\end{equation}
We observe that due to the orthonormality of the
decomposition basis, $R_{12}^{\alpha }=
\tilde{\Omega}/2$ and $S_{12}^{LR}=T_{21}^{RL}=0$. From the symmetry between
the two wells in the DWP structure it also follows that $E_{n}^{L}=E_{n}^{R}%
\equiv E_{n}$, $U_{n}^{L}=U_{n}^{R}\equiv U_{n}$, $U_{12}^{L}=U_{12}^{R}%
\equiv U_{12}$.
Using Eq. (\ref{canonical}), we derive explicit evolution equations
for the
Eq. (\ref{odesztheta}) and the evolution equations for the total numbers of
particles of each component $N_{n}$ and the respective phases,
$\gamma_{\alpha}=\theta_1^{\alpha}-\theta_2^{\alpha}$ (recall that
$\alpha=L,R$):
%\begin{widetext}
\begin{eqnarray}
\dot{z}_{n} &=&-\frac{2(K_{n}-K_{c,i}N_{n})}{\hbar }\,\sqrt{1-z_{n}^{2}}%
\,\sin \theta _{n}+\frac{V_{n}N_{n}}{2\hbar }(1-z_{n}^{2})\sin 2\theta _{n}
\nonumber  \label{odeszthetaapp} \\
&+&\frac{2}{\hbar }(V_{12}\sqrt{1-z_{3-n}^{2}}\,\cos \theta
_{3-n}+K_{c,12})N_{3-n}\sqrt{1-z_{n}^{2}}\,\sin \theta _{n}  \nonumber \\
&&\mp \frac{\tilde \Omega }{2\hbar }\sqrt{N_{1}N_{2}}
\bigg(\sqrt{(1+z_{1})(1+z_{2})}%
\sin \gamma _{L}-\sqrt{(1-z_{1})(1-z_{2})}\sin \gamma _{R}\bigg),
\nonumber
\\
\dot{\theta}_{n} &=&\frac{U_{n}-V_{n}}{\hbar }N_{n}z_{n}+\frac{%
2(K_{n}-K_{c,n}N_{n})}{\hbar }\,\frac{z_{n}\cos \theta _{n}}{\sqrt{%
1-z_{n}^{2}}}\nonumber\\
&-&\frac{V_{n}N_{n}}{2\hbar }\,z_{n}\cos 2\theta _{n}+\frac{%
U_{12}-V_{12}}{\hbar }N_{3-n}z_{3-n}  \nonumber \\
&-&\frac{2}{\hbar }\left[ V_{12}\sqrt{1-z_{3-n}^{2}}\,\cos \theta
_{3-n}+K_{c,12}\right] N_{3-n}\frac{z_{n}\cos \theta _{n}}
{\sqrt{1-z_{n}^{2}}%
}  \nonumber \\
&+&\frac{\tilde{\Omega}}{2\hbar }\sqrt{\frac{N_{3-n}}{N_{n}}}
\bigg(\sqrt{\frac{%
1+z_{3-n}}{1+z_{n}}}\cos \gamma _{L}-\sqrt{\frac{1-z_{3-n}}{1-z_{n}}}\cos
\gamma _{R}\bigg)\;.
\end{eqnarray}%
%\end{widetext}

%\begin{widetext}
\begin{eqnarray}
&&\dot{N_{n}}=\pm (-\frac{\tilde{\Omega}}{2\hbar }\sqrt{N_{1}N_{2}})\big(%
\sqrt{(1+z_{1})(1+z_{2})}\,\sin \gamma _{L}+\sqrt{(1-z_{1})(1-z_{2})}\,\sin
\gamma _{R}\big),
\nonumber
\label{nitetaalphaapp}
\\
&&\dot{\gamma}_{L}=\frac{1}{2\hbar }\bigg(%
N_{1}(U_{12}-U)(1+z_{1})-N_{2}(U_{12}-U)(1+z_{2})+\Delta E\bigg)\nonumber\\
&+&\frac{1}{%
\hbar }\bigg(K_{1}\sqrt{\frac{1-z_{1}}{1+z_{1}}}\cos \theta _{1}-K_{2}\sqrt{\frac{%
1-z_{2}}{1+z_{2}}}\cos \theta _{2}\bigg)
\nonumber
\\
&-&\frac{\tilde{\Omega}}{2\hbar \sqrt{N_{1}N_{2}}}\bigg(\frac{%
N_{2}(1+z_{2})-N_{1}(1+z_{1})}{\sqrt{(1+z_{1})(1+z_{2})}}\bigg)\cos \gamma
_{L}\nonumber\\
&+&\frac{K_{c,12}}{\hbar }\big[N_{1}\sqrt{1-z_{1}^{2}}\cos \theta
_{1}-N_{2}\sqrt{1-z_{2}^{2}}\cos \theta _{2}\big]-
\nonumber
\\
&&\frac{V}{2\hbar }\bigg[N_{1}(1-z_{1})(2+\cos 2\theta
_{1})-N_{2}(1-z_{2})(2+\cos 2\theta _{2})\bigg]-  \nonumber \\
&&\frac{1}{\hbar }\bigg[K_{c,1}N_{1}(2+z_{1})+K_{c,12}N_{2}+V_{12}N_{2}\sqrt{%
1-z_{2}^{2}}\cos \theta _{2}\bigg]\sqrt{\frac{1-z_{1}}{1+z_{1}}}\cos \theta
_{1}+
\nonumber
\\
&&\frac{1}{\hbar }\bigg[K_{c,2}N_{2}(2+z_{2})+K_{c,12}N_{1}\nonumber\\
&+&V_{12}N_{1}\sqrt{%
1-z_{1}^{2}}\cos \theta _{1}\bigg]\sqrt{\frac{1-z_{2}}{1+z_{2}}}\cos \theta
_{2}+\frac{V_{12}}{2}\big(N_{1}(1-z_{1})-N_{2}(1-z_{2})\big),
\nonumber
\\
&&\dot{\gamma}_{R}=\frac{1}{2\hbar }\bigg(%
N_{1}(U_{12}-U)(1-z_{1})-N_{2}(U_{12}-U)(1-z_{2})+\Delta E\bigg)\nonumber\\
&+&\frac{1}{%
\hbar }\bigg(K_{1}\sqrt{\frac{1+z_{1}}{1-z_{1}}}\cos \theta _{1}-K_{2}\sqrt{\frac{%
1+z_{2}}{1-z_{2}}}\cos \theta _{2}\bigg)-
\nonumber
\\
&&\frac{\tilde{\Omega}}{\hbar \sqrt{N_{1}N_{2}}}\bigg(\frac{%
N_{2}(1-z_{2})-N_{1}(1-z_{1})}{\sqrt{(1-z_{1})(1-z_{2})}}\bigg)\cos \gamma
_{R}\nonumber\\
&+&\frac{K_{c,12}}{\hbar }\big[N_{1}\sqrt{1-z_{1}^{2}}\cos \theta
_{1}-N_{2}\sqrt{1-z_{2}^{2}}\cos \theta _{2}\big]-
\nonumber
\\
&&\frac{V}{2\hbar }\bigg[N_{1}(1+z_{1})(2+\cos 2\theta
_{1})-N_{2}(1+z_{2})(2+\cos 2\theta _{2})\bigg]-  \nonumber \\
&&\frac{1}{\hbar }\bigg[K_{c,1}N_{1}(2-z_{1})+K_{c,12}N_{2}+V_{12}N_{2}\sqrt{%
1-z_{2}^{2}}\cos \theta _{2}\bigg]\sqrt{\frac{1+z_{1}}{1-z_{1}}}\cos \theta
_{1}+
\nonumber
\\
&&\frac{1}{\hbar }\bigg[K_{c,2}N_{2}(2-z_{2})+K_{c,12}N_{1}\nonumber\\
&+&V_{12}N_{1}\sqrt{%
1-z_{1}^{2}}\cos \theta _{1}\bigg]\sqrt{\frac{1+z_{2}}{1-z_{2}}}\cos \theta
_{2}+\frac{V_{12}}{2}\big(N_{1}(1+z_{1})-N_{2}(1+z_{2})\big),
\nonumber \\
&&
\end{eqnarray}
%\end{widetext}

\end{document}